\documentclass[sigconf, screen]{acmart}
\AtBeginDocument{%
  \providecommand\BibTeX{{%
    \normalfont B\kern-0.5em{\scshape i\kern-0.25em b}\kern-0.8em\TeX}}}

\copyrightyear{2024}
\acmYear{2024}
\setcopyright{rightsretained}
\acmConference[CHI EA '24]{Extended Abstracts of the CHI Conference on Human Factors in Computing Systems}{May 11--16, 2024}{Honolulu, HI, USA}
\acmBooktitle{Extended Abstracts of the CHI Conference on Human Factors in Computing Systems (CHI EA '24), May 11--16, 2024, Honolulu, HI, USA}
\acmDOI{10.1145/3613905.3650848}
\acmISBN{979-8-4007-0331-7/24/05}

\makeatletter

\newcommand\xlabel[2][]{\phantomsection\def\@currentlabelname{#1}\label{#2}}
\def\@fnsymbol#1{\ensuremath{\ifcase#1\or \text{\Letter} \or \medcircle\or
   \medtriangledown\or \medtriangleup\or \|\or \textbf{\ddag} \or \dagger\dagger
   \or \ddagger\ddagger \else\@ctrerr\fi}}

\makeatother

\usepackage{acmart-taps}
\usepackage{framed}
\usepackage{fancybox}
\usepackage{mathtools}
\usepackage{wrapfig}
\usepackage{longtable}
\usepackage{makecell}

\usepackage{subcaption}
\usepackage{enumitem}

\usepackage{booktabs}
\usepackage{colortbl}

\definecolor{redOV}{RGB}{255, 235, 238}
\definecolor{redI}{RGB}{255, 205, 210}
\definecolor{redII}{RGB}{239, 154, 154}
\definecolor{redIII}{RGB}{229, 115, 115}
\definecolor{redIV}{RGB}{239, 83, 80}
\definecolor{redV}{RGB}{244, 67, 54}
\definecolor{redVI}{RGB}{229, 57, 53}
\definecolor{redVII}{RGB}{211, 47, 47}
\definecolor{redVIII}{RGB}{198, 40, 40}
\definecolor{redIX}{RGB}{183, 28, 28}
\definecolor{redAI}{RGB}{255, 138, 128}
\definecolor{redAII}{RGB}{255, 82, 82}
\definecolor{redAIV}{RGB}{255, 23, 68}
\definecolor{redAVII}{RGB}{213, 0, 0}

\definecolor{pinkOV}{RGB}{252, 228, 236}
\definecolor{pinkI}{RGB}{248, 187, 208}
\definecolor{pinkII}{RGB}{244, 143, 177}
\definecolor{pinkIII}{RGB}{240, 98, 146}
\definecolor{pinkIV}{RGB}{236, 64, 122}
\definecolor{pinkV}{RGB}{233, 30, 99}
\definecolor{pinkVI}{RGB}{216, 27, 96}
\definecolor{pinkVII}{RGB}{194, 24, 91}
\definecolor{pinkVIII}{RGB}{173, 20, 87}
\definecolor{pinkIX}{RGB}{136, 14, 79}
\definecolor{pinkAI}{RGB}{255, 128, 171}
\definecolor{pinkAII}{RGB}{255, 64, 129}
\definecolor{pinkAIV}{RGB}{245, 0, 87}
\definecolor{pinkAVII}{RGB}{197, 17, 98}

\definecolor{purpleOV}{RGB}{243, 229, 245}
\definecolor{purpleI}{RGB}{225, 190, 231}
\definecolor{purpleII}{RGB}{206, 147, 216}
\definecolor{purpleIII}{RGB}{186, 104, 200}
\definecolor{purpleIV}{RGB}{171, 71, 188}
\definecolor{purpleV}{RGB}{156, 39, 176}
\definecolor{purpleVI}{RGB}{142, 36, 170}
\definecolor{purpleVII}{RGB}{123, 31, 162}
\definecolor{purpleVIII}{RGB}{106, 27, 154}
\definecolor{purpleIX}{RGB}{74, 20, 140}
\definecolor{purpleAI}{RGB}{234, 128, 252}
\definecolor{purpleAII}{RGB}{224, 64, 251}
\definecolor{purpleAIV}{RGB}{213, 0, 249}
\definecolor{purpleAVII}{RGB}{170, 0, 255}

\definecolor{deeppurpleOV}{RGB}{237, 231, 246}
\definecolor{deeppurpleI}{RGB}{209, 196, 233}
\definecolor{deeppurpleII}{RGB}{179, 157, 219}
\definecolor{deeppurpleIII}{RGB}{149, 117, 205}
\definecolor{deeppurpleIV}{RGB}{126, 87, 194}
\definecolor{deeppurpleV}{RGB}{103, 58, 183}
\definecolor{deeppurpleVI}{RGB}{94, 53, 177}
\definecolor{deeppurpleVII}{RGB}{81, 45, 168}
\definecolor{deeppurpleVIII}{RGB}{69, 39, 160}
\definecolor{deeppurpleIX}{RGB}{49, 27, 146}
\definecolor{deeppurpleAI}{RGB}{179, 136, 255}
\definecolor{deeppurpleAII}{RGB}{124, 77, 255}
\definecolor{deeppurpleAIV}{RGB}{101, 31, 255}
\definecolor{deeppurpleAVII}{RGB}{98, 0, 234}

\definecolor{indigoOV}{RGB}{232, 234, 246}
\definecolor{indigoI}{RGB}{197, 202, 233}
\definecolor{indigoII}{RGB}{159, 168, 218}
\definecolor{indigoIII}{RGB}{121, 134, 203}
\definecolor{indigoIV}{RGB}{92, 107, 192}
\definecolor{indigoV}{RGB}{63, 81, 181}
\definecolor{indigoVI}{RGB}{57, 73, 171}
\definecolor{indigoVII}{RGB}{48, 63, 159}
\definecolor{indigoVIII}{RGB}{40, 53, 147}
\definecolor{indigoIX}{RGB}{26, 35, 126}
\definecolor{indigoAI}{RGB}{140, 158, 255}
\definecolor{indigoAII}{RGB}{83, 109, 254}
\definecolor{indigoAIV}{RGB}{61, 90, 254}
\definecolor{indigoAVII}{RGB}{48, 79, 254}

\definecolor{blueOV}{RGB}{227, 242, 253}
\definecolor{blueI}{RGB}{187, 222, 251}
\definecolor{blueII}{RGB}{144, 202, 249}
\definecolor{blueIII}{RGB}{100, 181, 246}
\definecolor{blueIV}{RGB}{66, 165, 245}
\definecolor{blueV}{RGB}{33, 150, 243}
\definecolor{blueVI}{RGB}{30, 136, 229}
\definecolor{blueVII}{RGB}{25, 118, 210}
\definecolor{blueVIII}{RGB}{21, 101, 192}
\definecolor{blueIX}{RGB}{13, 71, 161}
\definecolor{blueAI}{RGB}{130, 177, 255}
\definecolor{blueAII}{RGB}{68, 138, 255}
\definecolor{blueAIV}{RGB}{41, 121, 255}
\definecolor{blueAVII}{RGB}{41, 98, 255}

\definecolor{lightblueOV}{RGB}{225, 245, 254}
\definecolor{lightblueI}{RGB}{179, 229, 252}
\definecolor{lightblueII}{RGB}{129, 212, 250}
\definecolor{lightblueIII}{RGB}{79, 195, 247}
\definecolor{lightblueIV}{RGB}{41, 182, 246}
\definecolor{lightblueV}{RGB}{3, 169, 244}
\definecolor{lightblueVI}{RGB}{3, 155, 229}
\definecolor{lightblueVII}{RGB}{2, 136, 209}
\definecolor{lightblueVIII}{RGB}{2, 119, 189}
\definecolor{lightblueIX}{RGB}{1, 87, 155}
\definecolor{lightblueAI}{RGB}{128, 216, 255}
\definecolor{lightblueAII}{RGB}{64, 196, 255}
\definecolor{lightblueAIV}{RGB}{0, 176, 255}
\definecolor{lightblueAVII}{RGB}{0, 145, 234}

\definecolor{cyanOV}{RGB}{224, 247, 250}
\definecolor{cyanI}{RGB}{178, 235, 242}
\definecolor{cyanII}{RGB}{128, 222, 234}
\definecolor{cyanIII}{RGB}{77, 208, 225}
\definecolor{cyanIV}{RGB}{38, 198, 218}
\definecolor{cyanV}{RGB}{0, 188, 212}
\definecolor{cyanVI}{RGB}{0, 172, 193}
\definecolor{cyanVII}{RGB}{0, 151, 167}
\definecolor{cyanVIII}{RGB}{0, 131, 143}
\definecolor{cyanIX}{RGB}{0, 96, 100}
\definecolor{cyanAI}{RGB}{132, 255, 255}
\definecolor{cyanAII}{RGB}{24, 255, 255}
\definecolor{cyanAIV}{RGB}{0, 229, 255}
\definecolor{cyanAVII}{RGB}{0, 184, 212}

\definecolor{tealOV}{RGB}{224, 242, 241}
\definecolor{tealI}{RGB}{178, 223, 219}
\definecolor{tealII}{RGB}{128, 203, 196}
\definecolor{tealIII}{RGB}{77, 182, 172}
\definecolor{tealIV}{RGB}{38, 166, 154}
\definecolor{tealV}{RGB}{0, 150, 136}
\definecolor{tealVI}{RGB}{0, 137, 123}
\definecolor{tealVII}{RGB}{0, 121, 107}
\definecolor{tealVIII}{RGB}{0, 105, 92}
\definecolor{tealIX}{RGB}{0, 77, 64}
\definecolor{tealAI}{RGB}{167, 255, 235}
\definecolor{tealAII}{RGB}{100, 255, 218}
\definecolor{tealAIV}{RGB}{29, 233, 182}
\definecolor{tealAVII}{RGB}{0, 191, 165}

\definecolor{greenOV}{RGB}{232, 245, 233}
\definecolor{greenI}{RGB}{200, 230, 201}
\definecolor{greenII}{RGB}{165, 214, 167}
\definecolor{greenIII}{RGB}{129, 199, 132}
\definecolor{greenIV}{RGB}{102, 187, 106}
\definecolor{greenV}{RGB}{76, 175, 80}
\definecolor{greenVI}{RGB}{67, 160, 71}
\definecolor{greenVII}{RGB}{56, 142, 60}
\definecolor{greenVIII}{RGB}{46, 125, 50}
\definecolor{greenIX}{RGB}{27, 94, 32}
\definecolor{greenAI}{RGB}{185, 246, 202}
\definecolor{greenAII}{RGB}{105, 240, 174}
\definecolor{greenAIV}{RGB}{0, 230, 118}
\definecolor{greenAVII}{RGB}{0, 200, 83}

\definecolor{lightgreenOV}{RGB}{241, 248, 233}
\definecolor{lightgreenI}{RGB}{220, 237, 200}
\definecolor{lightgreenII}{RGB}{197, 225, 165}
\definecolor{lightgreenIII}{RGB}{174, 213, 129}
\definecolor{lightgreenIV}{RGB}{156, 204, 101}
\definecolor{lightgreenV}{RGB}{139, 195, 74}
\definecolor{lightgreenVI}{RGB}{124, 179, 66}
\definecolor{lightgreenVII}{RGB}{104, 159, 56}
\definecolor{lightgreenVIII}{RGB}{85, 139, 47}
\definecolor{lightgreenIX}{RGB}{51, 105, 30}
\definecolor{lightgreenAI}{RGB}{204, 255, 144}
\definecolor{lightgreenAII}{RGB}{178, 255, 89}
\definecolor{lightgreenAIV}{RGB}{118, 255, 3}
\definecolor{lightgreenAVII}{RGB}{100, 221, 23}

\definecolor{limeOV}{RGB}{249, 251, 231}
\definecolor{limeI}{RGB}{240, 244, 195}
\definecolor{limeII}{RGB}{230, 238, 156}
\definecolor{limeIII}{RGB}{220, 231, 117}
\definecolor{limeIV}{RGB}{212, 225, 87}
\definecolor{limeV}{RGB}{205, 220, 57}
\definecolor{limeVI}{RGB}{192, 202, 51}
\definecolor{limeVII}{RGB}{175, 180, 43}
\definecolor{limeVIII}{RGB}{158, 157, 36}
\definecolor{limeIX}{RGB}{130, 119, 23}
\definecolor{limeAI}{RGB}{244, 255, 129}
\definecolor{limeAII}{RGB}{238, 255, 65}
\definecolor{limeAIV}{RGB}{198, 255, 0}
\definecolor{limeAVII}{RGB}{174, 234, 0}

\definecolor{yellowOV}{RGB}{255, 253, 231}
\definecolor{yellowI}{RGB}{255, 249, 196}
\definecolor{yellowII}{RGB}{255, 245, 157}
\definecolor{yellowIII}{RGB}{255, 241, 118}
\definecolor{yellowIV}{RGB}{255, 238, 88}
\definecolor{yellowV}{RGB}{255, 235, 59}
\definecolor{yellowVI}{RGB}{253, 216, 53}
\definecolor{yellowVII}{RGB}{251, 192, 45}
\definecolor{yellowVIII}{RGB}{249, 168, 37}
\definecolor{yellowIX}{RGB}{245, 127, 23}
\definecolor{yellowAI}{RGB}{255, 255, 141}
\definecolor{yellowAII}{RGB}{255, 255, 0}
\definecolor{yellowAIV}{RGB}{255, 234, 0}
\definecolor{yellowAVII}{RGB}{255, 214, 0}

\definecolor{amberOV}{RGB}{255, 248, 225}
\definecolor{amberI}{RGB}{255, 236, 179}
\definecolor{amberII}{RGB}{255, 224, 130}
\definecolor{amberIII}{RGB}{255, 213, 79}
\definecolor{amberIV}{RGB}{255, 202, 40}
\definecolor{amberV}{RGB}{255, 193, 7}
\definecolor{amberVI}{RGB}{255, 179, 0}
\definecolor{amberVII}{RGB}{255, 160, 0}
\definecolor{amberVIII}{RGB}{255, 143, 0}
\definecolor{amberIX}{RGB}{255, 111, 0}
\definecolor{amberAI}{RGB}{255, 229, 127}
\definecolor{amberAII}{RGB}{255, 215, 64}
\definecolor{amberAIV}{RGB}{255, 196, 0}
\definecolor{amberAVII}{RGB}{255, 171, 0}

\definecolor{orangeOV}{RGB}{255, 243, 224}
\definecolor{orangeI}{RGB}{255, 224, 178}
\definecolor{orangeII}{RGB}{255, 204, 128}
\definecolor{orangeIII}{RGB}{255, 183, 77}
\definecolor{orangeIV}{RGB}{255, 167, 38}
\definecolor{orangeV}{RGB}{255, 152, 0}
\definecolor{orangeVI}{RGB}{251, 140, 0}
\definecolor{orangeVII}{RGB}{245, 124, 0}
\definecolor{orangeVIII}{RGB}{239, 108, 0}
\definecolor{orangeIX}{RGB}{230, 81, 0}
\definecolor{orangeAI}{RGB}{255, 209, 128}
\definecolor{orangeAII}{RGB}{255, 171, 64}
\definecolor{orangeAIV}{RGB}{255, 145, 0}
\definecolor{orangeAVII}{RGB}{255, 109, 0}

\definecolor{deeporangeOV}{RGB}{251, 233, 231}
\definecolor{deeporangeI}{RGB}{255, 204, 188}
\definecolor{deeporangeII}{RGB}{255, 171, 145}
\definecolor{deeporangeIII}{RGB}{255, 138, 101}
\definecolor{deeporangeIV}{RGB}{255, 112, 67}
\definecolor{deeporangeV}{RGB}{255, 87, 34}
\definecolor{deeporangeVI}{RGB}{244, 81, 30}
\definecolor{deeporangeVII}{RGB}{230, 74, 25}
\definecolor{deeporangeVIII}{RGB}{216, 67, 21}
\definecolor{deeporangeIX}{RGB}{191, 54, 12}
\definecolor{deeporangeAI}{RGB}{255, 158, 128}
\definecolor{deeporangeAII}{RGB}{255, 110, 64}
\definecolor{deeporangeAIV}{RGB}{255, 61, 0}
\definecolor{deeporangeAVII}{RGB}{221, 44, 0}

\definecolor{brownOV}{RGB}{239, 235, 233}
\definecolor{brownI}{RGB}{215, 204, 200}
\definecolor{brownII}{RGB}{188, 170, 164}
\definecolor{brownIII}{RGB}{161, 136, 127}
\definecolor{brownIV}{RGB}{141, 110, 99}
\definecolor{brownV}{RGB}{121, 85, 72}
\definecolor{brownVI}{RGB}{109, 76, 65}
\definecolor{brownVII}{RGB}{93, 64, 55}
\definecolor{brownVIII}{RGB}{78, 52, 46}
\definecolor{brownIX}{RGB}{62, 39, 35}

\definecolor{grayOV}{RGB}{250, 250, 250}
\definecolor{grayI}{RGB}{245, 245, 245}
\definecolor{grayII}{RGB}{238, 238, 238}
\definecolor{grayIII}{RGB}{224, 224, 224}
\definecolor{grayIV}{RGB}{189, 189, 189}
\definecolor{grayV}{RGB}{158, 158, 158}
\definecolor{grayVI}{RGB}{117, 117, 117}
\definecolor{grayVII}{RGB}{97, 97, 97}
\definecolor{grayVIII}{RGB}{66, 66, 66}
\definecolor{grayIX}{RGB}{33, 33, 33}

\definecolor{bluegrayOV}{RGB}{236, 239, 241}
\definecolor{bluegrayI}{RGB}{207, 216, 220}
\definecolor{bluegrayII}{RGB}{176, 190, 197}
\definecolor{bluegrayIII}{RGB}{144, 164, 174}
\definecolor{bluegrayIV}{RGB}{120, 144, 156}
\definecolor{bluegrayV}{RGB}{96, 125, 139}
\definecolor{bluegrayVI}{RGB}{84, 110, 122}
\definecolor{bluegrayVII}{RGB}{69, 90, 100}
\definecolor{bluegrayVIII}{RGB}{55, 71, 79}
\definecolor{bluegrayIX}{RGB}{38, 50, 56}
\definecolor{bluegrayX}{RGB}{17, 23, 26}

\definecolor{myACMBlue}{cmyk}{1,0.1,0,0.1}
\definecolor{myACMYellow}{cmyk}{0,0.16,1,0}
\definecolor{myACMOrange}{cmyk}{0,0.42,1,0.01}
\definecolor{myACMRed}{cmyk}{0,0.90,0.86,0}
\definecolor{myACMLightBlue}{cmyk}{0.49,0.01,0,0}
\definecolor{myACMGreen}{cmyk}{0.20,0,1,0.19}
\definecolor{myACMPurple}{cmyk}{0.55,1,0,0.15}
\definecolor{myACMDarkBlue}{cmyk}{1,0.58,0,0.21}
\definecolor{myReferenceURLColor}{HTML}{09326F}

\newcommand{\mylink}[1]{{\href{#1}{\color{blueVI}\textbf{\texttt{#1}}}}}

\definecolor{platformcolor}{HTML}{9D7660}
\definecolor{communicationcolor}{HTML}{59a14f}
\definecolor{datacolor}{HTML}{4e79a7}
\definecolor{displaycolor}{HTML}{F28E2D}
\definecolor{modularitycolor}{HTML}{e15759}

\newcommand{\platformc}[1]{\textcolor{platformcolor}{#1}}
\newcommand{\communicationc}[1]{\textcolor{communicationcolor}{#1}}
\newcommand{\datac}[1]{\textcolor{datacolor}{#1}}
\newcommand{\displayc}[1]{\textcolor{displaycolor}{#1}}
\newcommand{\modularityc}[1]{\textcolor{modularitycolor}{#1}}
\newcommand{\implementationc}[1]{\textcolor{grayVI}{#1}}

\newcommand{\mypar}[1]{\vspace{3pt}\noindent\textbf{#1}}

\newcommand{\figpart}[1]{\textcolor{myACMPurple}{#1}}
\hypersetup{
  colorlinks,
  linkcolor=blueVIII,
  citecolor=blueVIII,
  urlcolor=myReferenceURLColor,
  filecolor=myACMDarkBlue
}

\newcommand{\tool}{\textsc{SuperNOVA}}

\newcommand{\papercount}{64}

\newcommand{\packagecount}{105}

\newcommand{\totalcount}{163}
\newcommand{\totalpackagecount}{984}

\newcommand{\notebookcount}{8.6 million}
\newcommand{\vanotebookcount}{55k}

\newcommand{\package}[1]{\textsc{#1}}

\newcommand{\headertopspace}{\vspace{0pt}}
\newcommand{\headerbottomspace}{\vspace{0pt}}

\newcommand*{\vcenteredhbox}[1]{\begingroup\setbox0=\hbox{#1}\parbox{\wd0}{\box0}\endgroup}

\definecolor{tablerowcolor}{HTML}{EEEEEE}

\definecolor{boxbackground}{RGB}{245, 245, 245}
\definecolor{boxbackgrounduser}{RGB}{245, 245, 245}
\definecolor{boxbackgroundwhy}{RGB}{245, 245, 245}

\definecolor{boxborder}{HTML}{A0D8D3}
\definecolor{boxtext}{HTML}{14312F}

\newcommand{\columnbox}[1]{
  \setlength{\fboxrule}{0.8pt}
  \setlength{\fboxsep}{3pt}
  \vspace{3pt}
  \noindent\fcolorbox{grayIV}{white}{\parbox{\dimexpr\linewidth-2\fboxsep-2\fboxrule\relax}{#1}}
}

\begin{document}

\title{\tool{}: Design Strategies and Opportunities for Interactive Visualization in Computational Notebooks}

\author{Zijie J. Wang}
\orcid{0000-0003-4360-1423}
\affiliation{%
  \institution{Georgia Tech}
  \city{Atlanta}
  \state{Georgia}
  \country{USA}
}

\author{David Munechika}
\orcid{0000-0002-3643-6899}
\affiliation{%
  \institution{Georgia Tech}
  \city{Atlanta}
  \state{Georgia}
  \country{USA}
}

\author{Seongmin Lee}
\orcid{0000-0002-1950-5004}
\affiliation{%
  \institution{Georgia Tech}
  \city{Atlanta}
  \state{Georgia}
  \country{USA}
}

\author{Duen Horng Chau}
\orcid{0000-0001-9824-3323}
\affiliation{%
  \institution{Georgia Tech}
  \city{Atlanta}
  \state{Georgia}
  \country{USA}
}

\renewcommand{\shortauthors}{Zijie J. Wang, et al.}

\begin{abstract}
  Computational notebooks, such as Jupyter Notebook, have become data scientists' de facto programming environments.
  Many visualization researchers and practitioners have developed interactive visualization tools that support notebooks, yet little is known about the appropriate design of these tools.
  To address this critical research gap, we investigate the design strategies in this space by analyzing \totalcount{} notebook visualization tools.
  Our analysis encompasses \papercount{} systems from academic papers and \packagecount{} systems sourced from a pool of \vanotebookcount{} notebooks containing interactive visualizations that we obtain via scraping \notebookcount{} notebooks on GitHub.
  Through this study, we identify key design implications and trade-offs, such as leveraging multimodal data in notebooks as well as balancing the degree of visualization-notebook integration.
  Furthermore, we provide empirical evidence that tools compatible with more notebook platforms have a greater impact.
  Finally, we develop \tool{}, an open-source interactive browser to help researchers explore existing notebook visualization tools.
  \tool{} is publicly accessible at: \mylink{https://poloclub.github.io/supernova/}.
\end{abstract} 

\ccsdesc[500]{Human-centered computing~Visualization}
\ccsdesc[500]{Human-centered computing~Interactive systems and tools}
\ccsdesc[500]{Human-centered computing~Visualization systems and tools}
\ccsdesc[500]{Human-centered computing~Visualization design and evaluation methods}

\keywords{Computational Notebook, Interactive Visualization, Systematic Review, Data Science, Design, Cross-Platform Visualization}

\begin{teaserfigure}
  \includegraphics[width=0.97\linewidth]{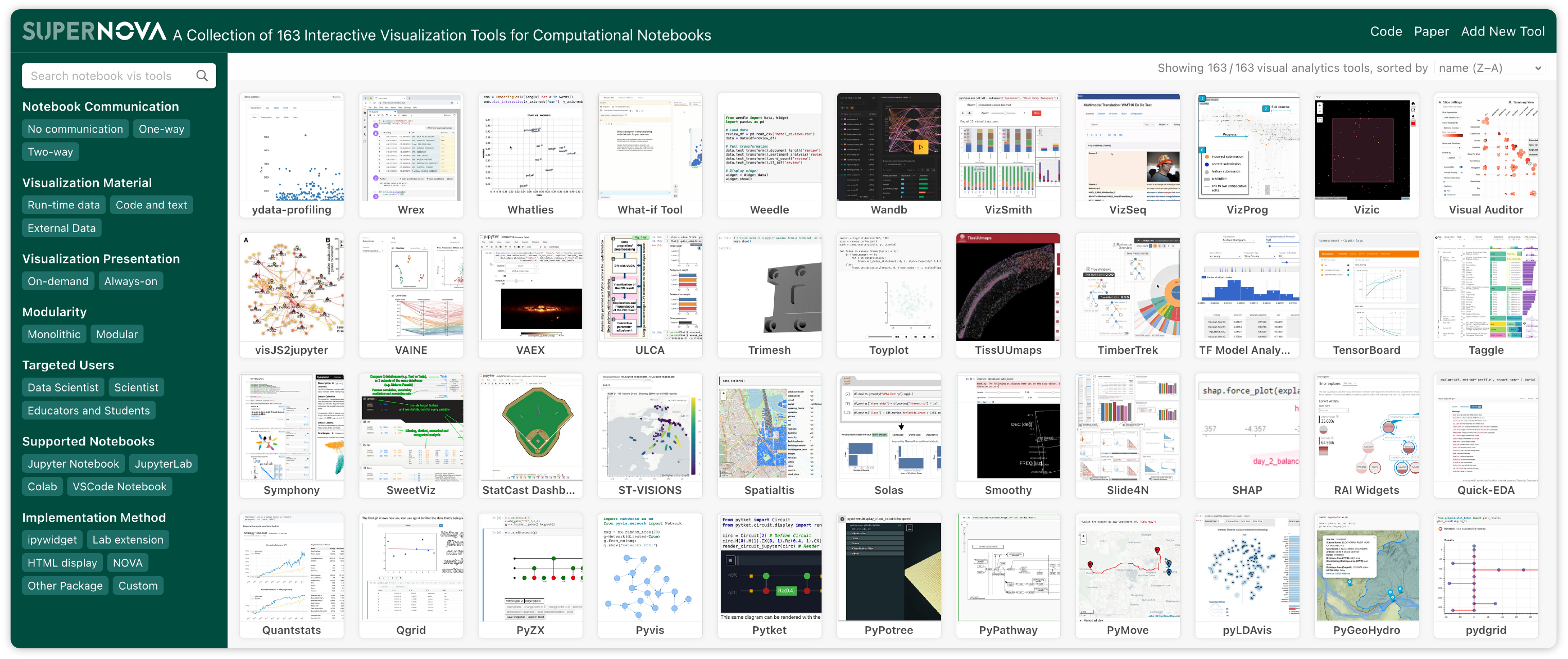}
  \vspace{-10pt}
  \caption{
    \tool{} is a browser for exploring \totalcount{} notebook interactive visualization tools.
    Users can filter and search for tools with specific properties in the left panel.
    Clicking on a tool reveals details including paper metadata and GitHub repository.
  }
  \label{fig:supernova}
\end{teaserfigure}

\maketitle

\section{Introduction}

Computational notebooks, such as Jupyter Notebook~\cite{kluyverJupyterNotebooksPublishing2016} and Colab, are the most popular programming environments among data scientists~\cite{kaggleStateMachineLearning2022}.
These notebooks seamlessly combine text, code, and visual outputs in a document that consists of an arbitrary number of \textit{cells}---small text and code editors.
Users can execute a code cell, and its output (e.g., text and visualizations) will be displayed below the cell.
By providing a literate programming environment, notebooks enable users to perform exploratory data analysis, document their work, and share insights with collaborators~\cite{ruleExplorationExplanationComputational2018}.

To create easy-to-adopt tools, there is a trend in the VIS community to develop interactive visualization systems that can be used in notebooks~\cite[e.g.,][]{onoPipelineProfilerVisualAnalytics2021,xenopoulosCalibrateInteractiveAnalysis2023,wangTimberTrekExploringCurating2022a}.
Designing visualizations for notebook environments presents unique opportunities and considerations.
On the one hand, notebook visualization tools allow direct modification of data through user interactions~\cite{uberDeckGlWebGL22016}, and users can mix-and-match different visualization tools to create dashboards~\cite{wangStickyLandBreakingLinear2022}.
However, notebook users often write fragmentary code
and execute it nonlinearly~\cite{mcnuttDesignAIpoweredCode2023, weinmanForkItSupporting2021}, which differs from traditional workflows for using interactive visualization systems~\cite{chenWhatMayVisualization2016}.

Therefore, if researchers do not consider notebooks' unique characteristics, their notebook visualization tools may not fully realize the potential of notebooks and, at worst, may impede the ability of notebook users to effectively use these tools.
To shed light on the existing landscape of notebook visualization tools and help visualization researchers and practitioners harness the potential of notebook environments, we \textbf{contribute:}

\begin{figure}[tb]
  \includegraphics[width=\linewidth]{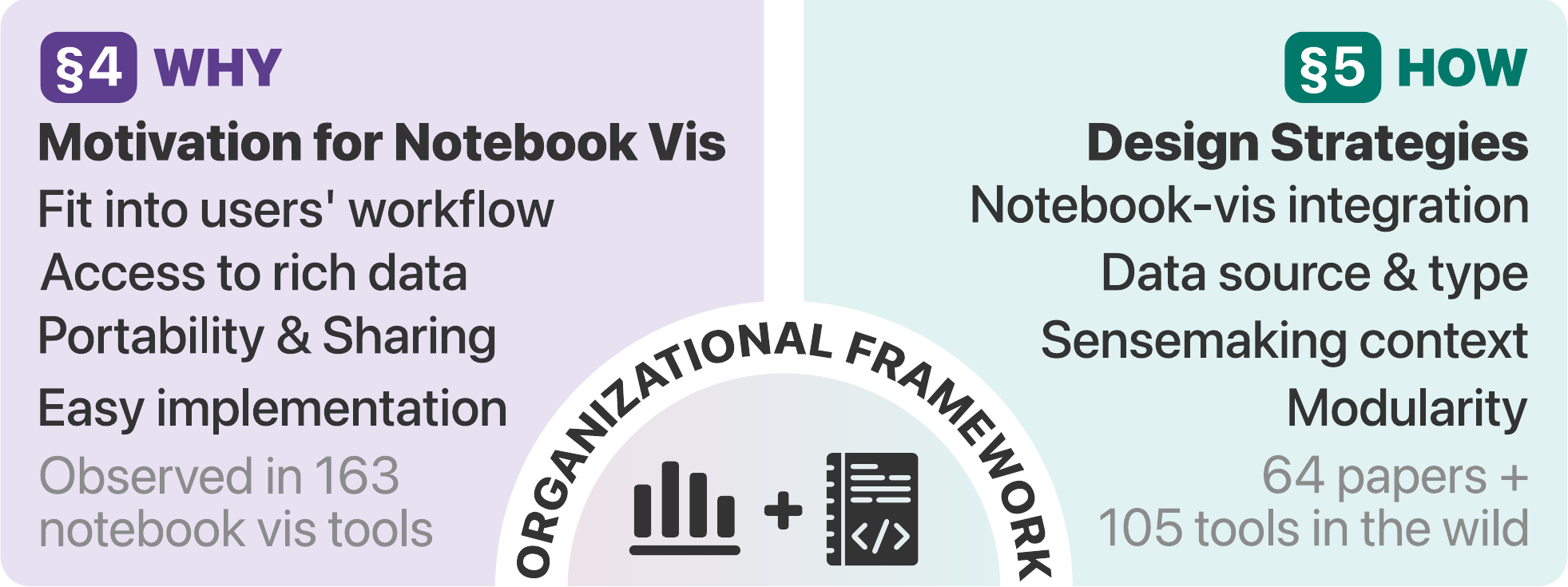}
  \caption{
    We present an organizational framework to characterize notebook visualization tools based on their design motivations and strategies through a review of \totalcount{} tools.
  }
  \label{fig:overview}
  \Description{This figure has two sections, labeled "Section 4 WHY" and "Section 5 HOW," each with a different background color, purple and teal respectively. The "Section 4 WHY" section includes the title "Motivation for notebook vis" and lists factors such as "Fit into users' workflow," "Access to rich data," "Portability & Sharing," "Easy implementation," and "Observed in 161 notebook vis tools." The "Section 5 HOW" section contains the title "Design strategies" and lists items like "Vis-notebook integration," "Data source & type," "Sensemaking context," and "Modularity," followed by "64 papers + 103 tools in the wild." At the bottom, there is an "ORGANIZATIONAL FRAMEWORK" bar spanning both sections, depicting icons for charts, a building, code brackets, and an integration symbol, suggesting a structure for organizing the information.}
\end{figure}

\aptLtoX[graphic=no,type=html]{
      \begin{itemize}
            \item \textbf{The first systematic review of \totalcount{} notebook visualization tools} including \papercount{} systems introduced in academic papers and \packagecount{} tools sourced from a pool of \vanotebookcount{} notebooks containing interactive visualizations that we obtain via scraping \notebookcount{} notebooks on GitHub~(\autoref{fig:overview}).
                  To inform the design of future tools, we discuss unique design implications~(\autoref{sec:why}) and trade-offs~(\autoref{sec:design}).

            \item \textbf{Organizational framework to characterize notebook visualization tools} in terms of their motivation for supporting notebooks~(\autoref{sec:why}), targeted users~(\autoref{sec:why:workflow}), and a four-dimensional design space based on user needs~(\autoref{sec:design}).
                  This framework facilitates a more comprehensive understanding of the landscape of notebook visualization tools.
                  Based on this framework, we further analyze the effects of design factors on the impact of notebook visualization tools.
                  We find tools supporting more notebook platforms have significantly more GitHub stars and paper citations~(\autoref{sec:evaluation}).
      \end{itemize}
}{
      \begin{itemize}[topsep=1pt, itemsep=0mm, parsep=2pt, leftmargin=8pt]
            \item \textbf{The first systematic review of \totalcount{} notebook visualization tools} including \papercount{} systems introduced in academic papers and \packagecount{} tools sourced from a pool of \vanotebookcount{} notebooks containing interactive visualizations that we obtain via scraping \notebookcount{} notebooks on GitHub~(\autoref{fig:overview}).
                  To inform the design of future tools, we discuss unique design implications~(\autoref{sec:why}) and trade-offs~(\autoref{sec:design}).

            \item \textbf{Organizational framework to characterize notebook visualization tools} in terms of their motivation for supporting notebooks~(\autoref{sec:why}), targeted users~(\autoref{sec:why:workflow}), and a four-dimensional design space based on user needs~(\autoref{sec:design}).
                  This framework facilitates a more comprehensive understanding of the landscape of notebook visualization tools.
                  Based on this framework, we further analyze the effects of design factors on the impact of notebook visualization tools.
                  We find tools supporting more notebook platforms have significantly more GitHub stars and paper citations~(\autoref{sec:evaluation}).
      \end{itemize}
}

\noindent{}To broaden the public's access to our collection, we develop \tool{}~(\autoref{fig:supernova}), an interactive tool that helps researchers and designers explore existing notebook visualization tools and search for design inspiration and implementation references.
Anyone can easily add new tools to this open-source\footnote{\tool{} code: \mylink{https://github.com/poloclub/supernova}} explorer.
\tool{} is publicly accessible at: \mylink{https://poloclub.github.io/supernova/}.

\headertopspace{}
\section{Related Work}
\headerbottomspace{}

Our work joins the research body of studying interactive tools for notebooks.
To understand notebook users' behaviors, researchers conduct interview studies~\cite{keryStoryNotebookExploratory2018} and analyze 1 million notebooks scraped from GitHub~\cite{ruleExplorationExplanationComputational2018}.
Researchers present methods to help researchers \textit{develop} notebook-compatible visualization tools~\cite{piazentinonoInteractiveDataVisualization2021,wangNOVAPracticalMethod2022a}.
More recently, a design space analysis is conducted for AI-powered code assistants for notebooks~\cite{mcnuttDesignAIpoweredCode2023}.
In contrast, our work focuses on the design of visualization tools for notebooks by analyzing \totalcount{} tools identified from academic papers and \notebookcount{} notebooks.
Additionally, inspired by the popular interactive survey browsers for text visualization~\cite{kucherTextVisualizationTechniques2015}, biological data visualization~\cite{kerrenBioVisExplorerVisual2017}, visualizations for trust in machine learning~\cite{chatzimparmpasVisualizationTrustMachine2024}, and embedding visualization~\cite{huangVAEmbeddingsSTAR2023}, we develop \tool{}, the first interactive explorer for notebook visualization tools. %
\headertopspace{}
\section{Methodology}
\label{sec:method}
\headerbottomspace{}

\mypar{Systematic Review.}
To study how researchers and practitioners design notebook visualization tools, we collected and analyzed \papercount{} academic papers and \packagecount{} tools in the wild.
In this study, we define notebook visualization tools as systems that can display interactive visualizations in Python computational notebooks.
\textbf{(1)~Literature collection:} we searched Google Scholar for notebook visualization tools and performed forward and backward reference searches to snowball the results.
\textbf{(2)~In-the-wild tool collection:} we scraped \notebookcount{} notebooks from GitHub and filtered \vanotebookcount{} notebooks containing interactive visualizations by matching notebook cell output types.
We extracted \totalpackagecount{} potential visualization packages by matching variable names and imported modules using abstract syntax trees, and we manually examined each package to keep \packagecount{} that were indeed notebook visualization tools (see \autoref{sec:appendix:collection} for details).
\textbf{(3)~Coding:} we conducted a multi-phase coding process to analyze the collected papers, documentation, and demo notebooks.
First, three authors independently open coded~\cite{braunUsingThematicAnalysis2006} the same 30 random tools regarding the motivations for using notebooks and design strategies using Google Sheets.
After discussing the codebook and resolving disagreements, the three coders independently conducted open coding on the remaining tools, allocating an equal number of tools to each author.
Following the analysis of the final codebook and themes, one author applied deductive coding~\cite{merriamIntroductionQualitativeResearch2002} to assign identified design patterns to each tool.
We share all scraping code, codebook, and metadata of \totalcount{} tools in \tool{}'s \href{https://github.com/supernova2023/supernova}{\color{blueVI}\textbf{repository}}.

\setlength{\abovecaptionskip}{5pt}
\setlength{\belowcaptionskip}{0pt}
\setlength{\abovecaptionskip}{10pt}

\mypar{Organizational Framework.}
Our large-scale systematic review resulted in an organizational framework characterizing notebook visualization tools in terms of motivations for supporting notebooks~(\autoref{sec:why}), targeted users~(\autoref{sec:why:workflow}), and design patterns based on user needs~(\autoref{sec:design}).
Using this framework, we develop \tool{}~(\autoref{fig:supernova}), an interactive explorer that allows for easy filtering and searching for notebook visualization tools with desired properties.
Based on our review, we distill 4 design implications and 4 design trade-offs to help future researchers design notebook visualization tools.
Finally, we conduct a correlation analysis and two regression analyses to examine the effects of design patterns on the impacts of notebook interactive visualization tools~(\autoref{sec:evaluation}). %
\headertopspace{}
\section{Why Notebook Visualization Tools}
\label{sec:why}
\headerbottomspace{}

This section discusses the motivation for developing interactive visualization tools for computational notebooks.
We organize these motivations into four non-mutually exclusive groups.

\headertopspace{}
\subsection{Seamless Workflow Integration}
\label{sec:why:workflow}
\headerbottomspace{}

\setlength{\abovecaptionskip}{5pt}
\begin{figure}[tb]
  \includegraphics[width=\linewidth]{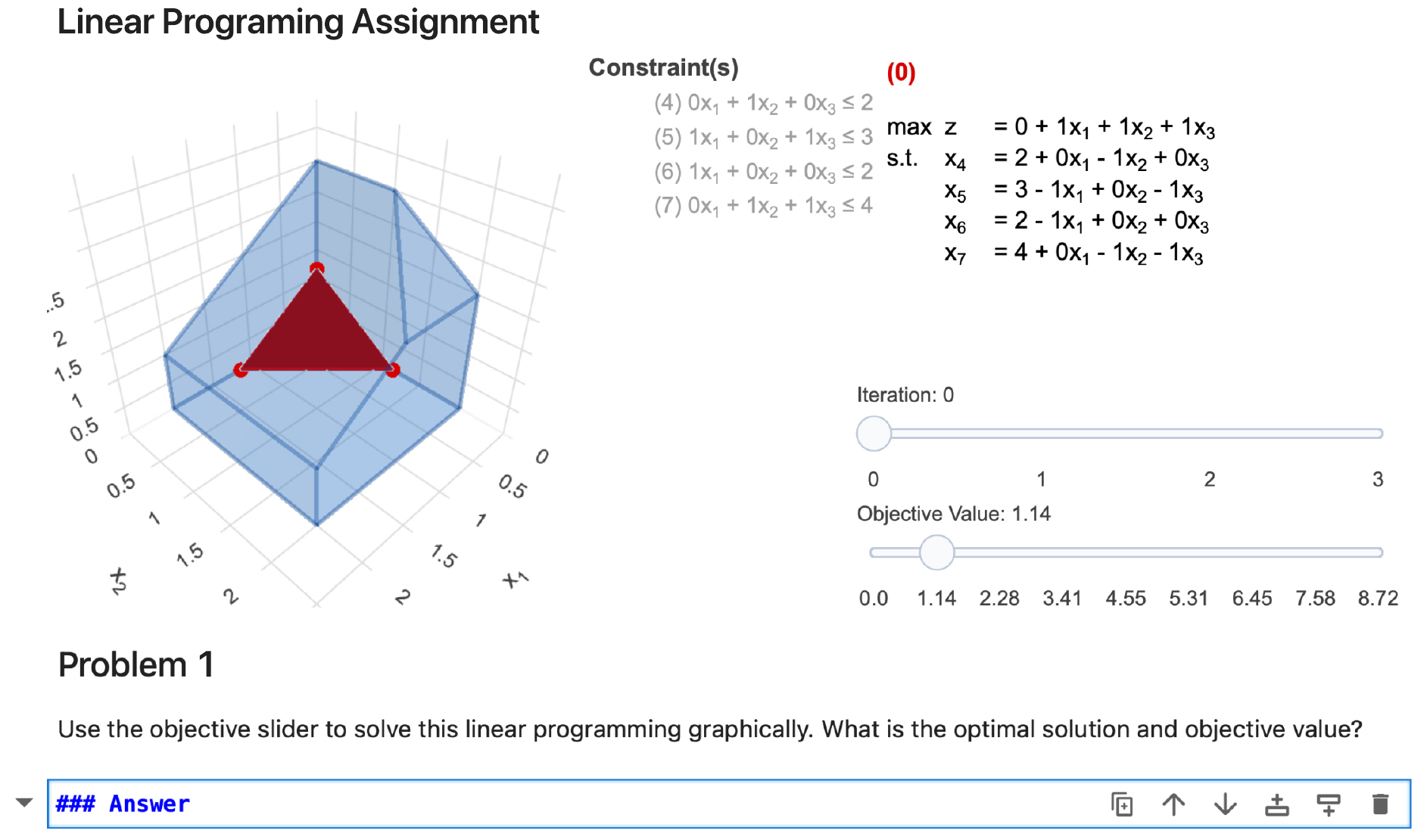}
  \caption{
    Many notebook visualization tools are developed for educators and students, such as GILP~\cite{robbinsGILPInteractiveTool2023} which offers interactive and easy-to-understand visualizations to help students learn about linear programming algorithms.
    Educators can directly integrate GILP into notebook-based assignments.
  }
  \label{fig:education}
  \Description{Screenshot of the GILP package.}
\end{figure}
\setlength{\belowcaptionskip}{0pt}
\setlength{\abovecaptionskip}{10pt}

Our study reveals that most of the surveyed visualization tools support notebooks as a means of aligning with the workflows of end-users.
We observe that different user groups have distinct notebook usage patterns.
Therefore, to ground our discussion on the notebook workflows of end-users, we categorize end-users into three user groups: data scientists, scientists, and educators and students.

\mypar{\vcenteredhbox{\includegraphics[height=8pt]{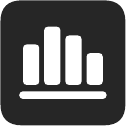}}~Data Scientists.}
Notebooks are the most popular programming environment among data scientists~\cite{kaggleStateMachineLearning2022}.
Consequently, many researchers have developed notebook visualization tools to promote adoption among data scientists.
Data scientists use notebooks for conducting rapid experiments, collaborating with other stakeholders, and directly deploying notebooks within production pipelines~\cite{chattopadhyayWhatWrongComputational2020}.
Notebook visualization tools have covered almost every stage of data scientists' workflow, from annotating data~\cite{zhangMEGAnnoExploratoryLabeling2023} and exploring data~\cite{liEDAssistantSupportingExploratory2023}, to developing ML models~\cite{onoPipelineProfilerVisualAnalytics2021}, documenting models~\cite{bhatAspirationsPracticeModel2023},  evaluating models~\cite{munechikaVisualAuditorInteractive2022}, and communicating findings to stakeholders~\cite{wangSlide4NCreatingPresentation2023}.

\mypar{\vcenteredhbox{\includegraphics[height=8pt]{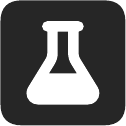}}~Scientists.}
Notebooks are also popular among scientists, including biologists and physicists.
Scientists use them as an interface for accessing remote clusters~\cite{sbailoNOMADArtificialIntelligenceToolkit2022}, and publishing notebooks with academic papers is considered good practice for reproducible research~\cite{herwigCyberhubsVirtualResearch2018}.
Thus, many notebook visualization tools are developed to facilitate scientific research workflows, such as designing experiments~\cite{guoVAINEVisualizationAI2021}, simulating physical environments~\cite{freemanBraxDifferentiablePhysics2021}, and analyzing molecules~\cite{nguyenNGLviewInteractiveMolecular2018} and astronomical data~\cite{arayaJOVIALNotebookbasedAstronomical2018}.

\mypar{\vcenteredhbox{\includegraphics[height=8pt]{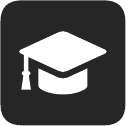}}~Educators and Students.}
Notebooks are increasingly being used as interactive textbooks in computing education, as they enable students to easily interact with code and test their ideas~\cite{smithModelingStudentEngagement2021}.
Educators also use notebooks for assigning and grading programming assignments~\cite{hullVISGRADERAutomaticGrading2023}.
In this use case, notebooks serve as worksheets where students write and run their code in specific cells.
We observe a growing trend of notebook visualization tools that are specifically developed for educators and students.
For example, \package{GILP}~\cite{robbinsGILPInteractiveTool2023} visualizes simplex algorithms in notebooks, allowing educators to design interactive textbooks and assignments~(\autoref{fig:education}).
\package{VizProg}~\cite{zhangVizProgIdentifyingMisunderstandings2023} helps instructors monitor students' coding progress during in-class exercises through interactive visualizations.

Our findings highlight that computational notebooks are a popular medium among diverse user groups.
In addition to data scientists, scientists, educators, and students also use notebooks in their workflows.
This provides visualization researchers and designers with exciting opportunities to develop tools that can be easily adopted.
However, we find different user groups have distinct notebook workflows.
For example, scientists use notebooks for collaboration and reproducible research, while educators use them as textbooks and worksheets.
Therefore, researchers should engage with targeted user groups in the early design process~\cite{sedlmairDesignStudyMethodology2012} to investigate users' notebook workflows and ground their designs.

\aptLtoX{\begin{framed}\noindent\textbf{Implication on domain-specific design}:
Designing notebook visualization tools requires researchers to engage with targeted user groups to develop tailored tools, as different user groups have distinct notebook usage patterns.
\end{framed}}
{\columnbox{
{\xlabel[Implecation: domain-specific design]{opp:user-group}}
  \textbf{Implication on domain-specific design}:
  Designing notebook visualization tools requires researchers to engage with targeted user groups to develop tailored tools, as different user groups have distinct notebook usage patterns.}}

\headertopspace{}
\subsection{Easy Access to Read and Refine Artifacts}
\label{sec:why:artifact}
\headerbottomspace{}

\setlength{\belowcaptionskip}{-10pt}
\setlength{\abovecaptionskip}{5pt}
\begin{figure}[tb]
  \includegraphics[width=\linewidth]{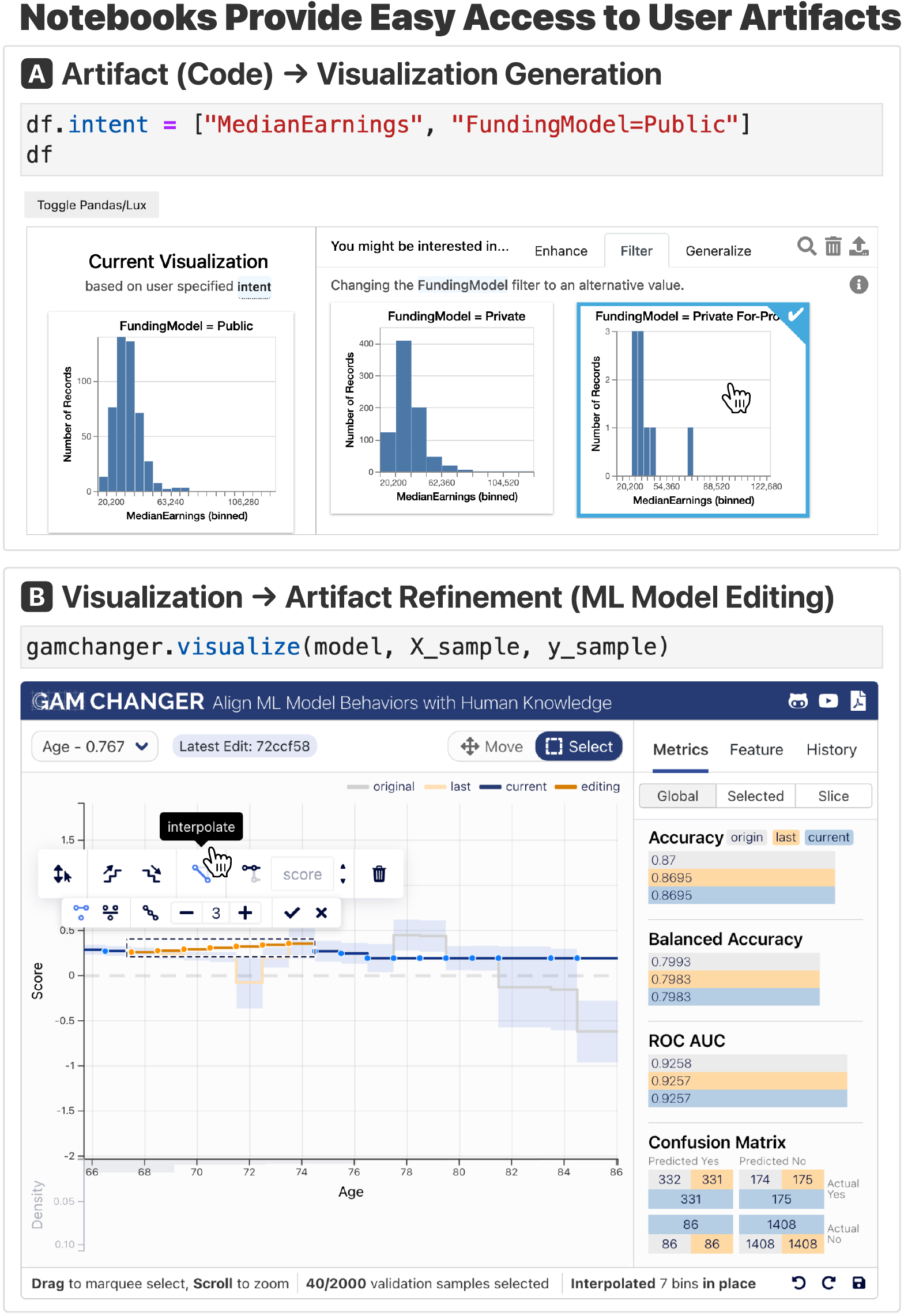}
  \caption{
    Computational notebooks offer unique opportunities for visualization tools to \textit{read} and \textit{refine} users' artifacts, such as code, data, and models.
    For example, (A) \package{Lux}~\cite{leeLuxAlwaysonVisualization2021} leverages a user's data transformation code to recommend visualizations, while (B) \package{GAM Changer}~\cite{wangInterpretabilityThenWhat2022a} enables users to interactively edit an ML model's learned weights.
  }
  \label{fig:artifact}
  \Description{Screenshots of Lux and GAM Changer packages.}
\end{figure}
\setlength{\belowcaptionskip}{0pt}
\setlength{\abovecaptionskip}{10pt}

Notebook visualization tools not only benefit from easy adoption but also access to programming artifacts, including code, raw data, and models.
These tools can be categorized into two groups based on their uses of artifacts.

\mypar{Artifacts $\rightarrow$ Visualization Generation.}
To create visualizations in non-notebook environments, data scientists often need to \textit{manually} specify chart types and input data.
However, notebook tools have access to all artifacts needed to create visualizations.
For example, \package{B2}~\cite{wuB2BridgingCode2020} uses dataframes and code queries in notebooks to \textit{automatically} synthesize interactive visualizations.
Similarly, \package{Lux}~\cite{leeLuxAlwaysonVisualization2021} and \package{Solas}~\cite{eppersonLeveragingAnalysisHistory2022} provide automatic visualization recommendations based on a user's dataframe and analysis history~(\autoref{fig:artifact}\figpart{A}).
Through accessing ML models that are being trained in notebooks, \package{TensorBoard}~\cite{abadiTensorFlowSystemLargescale2016} can visualize the model's performance in real time.

\mypar{Visualizations $\rightarrow$ Artifact Refinement.}
After gaining insights from visualizations, data scientists often \textit{manually} refine their code, data, and models outside of notebooks.
Notebooks can accelerate this process by directly updating artifacts.
For example, \package{Mage}~\cite{keryMageFluidMoves2020} automatically generates code to reflect the change caused by a user's interaction with visualizations (e.g., deleting a column from a table).
Similarly, \package{GAM Changer}~\cite{wangInterpretabilityThenWhat2022a} enables users to modify ML model weights by direct manipulation on visualizations~(\autoref{fig:artifact}\figpart{B}).

When designing notebook visualization tools, it is crucial to consider integrating the \textit{input} and \textit{output} in the visualization workflow~\cite[e.g.,][]{chenWhatMayVisualization2016, cashmanUserBasedVisual2019, upsonApplicationVisualizationSystem1989} into the notebook environment.
Take \citet{keimVisualAnalyticsDefinition2008}'s visual analytics pipeline as an example, the \textit{input data} can be notebook runtime artifacts, text, and usage logs~(\autoref{sec:design:data}), and the \textit{output knowledge} can be directly operationalized to synthesize code, transform data, and update ML models in the notebook~(\autoref{sec:design:communication}).

\aptLtoX{\begin{framed}\noindent\textbf{Implication on new opportunities enabled by easy artifact access}:
  Computational notebooks provide unique opportunities for researchers to integrate the \textit{input} of a visualization pipeline~(e.g., notebook runtime artifacts and text) and operationalize its \textit{output}~(e.g., transforming data and updating ML models) within the users' existing workflow.\end{framed}}{\columnbox{
{\xlabel[Implecation: new opportunities]{opp:artifact}}
\textbf{Implication on new opportunities enabled by easy artifact access}:
  Computational notebooks provide unique opportunities for researchers to integrate the \textit{input} of a visualization pipeline~(e.g., notebook runtime artifacts and text) and operationalize its \textit{output}~(e.g., transforming data and updating ML models) within the users' existing workflow.}}

\subsection{Portability and Shareability}
\label{sec:why:portability}
\headerbottomspace{}

\begin{figure}[tb]
  \includegraphics[width=\linewidth]{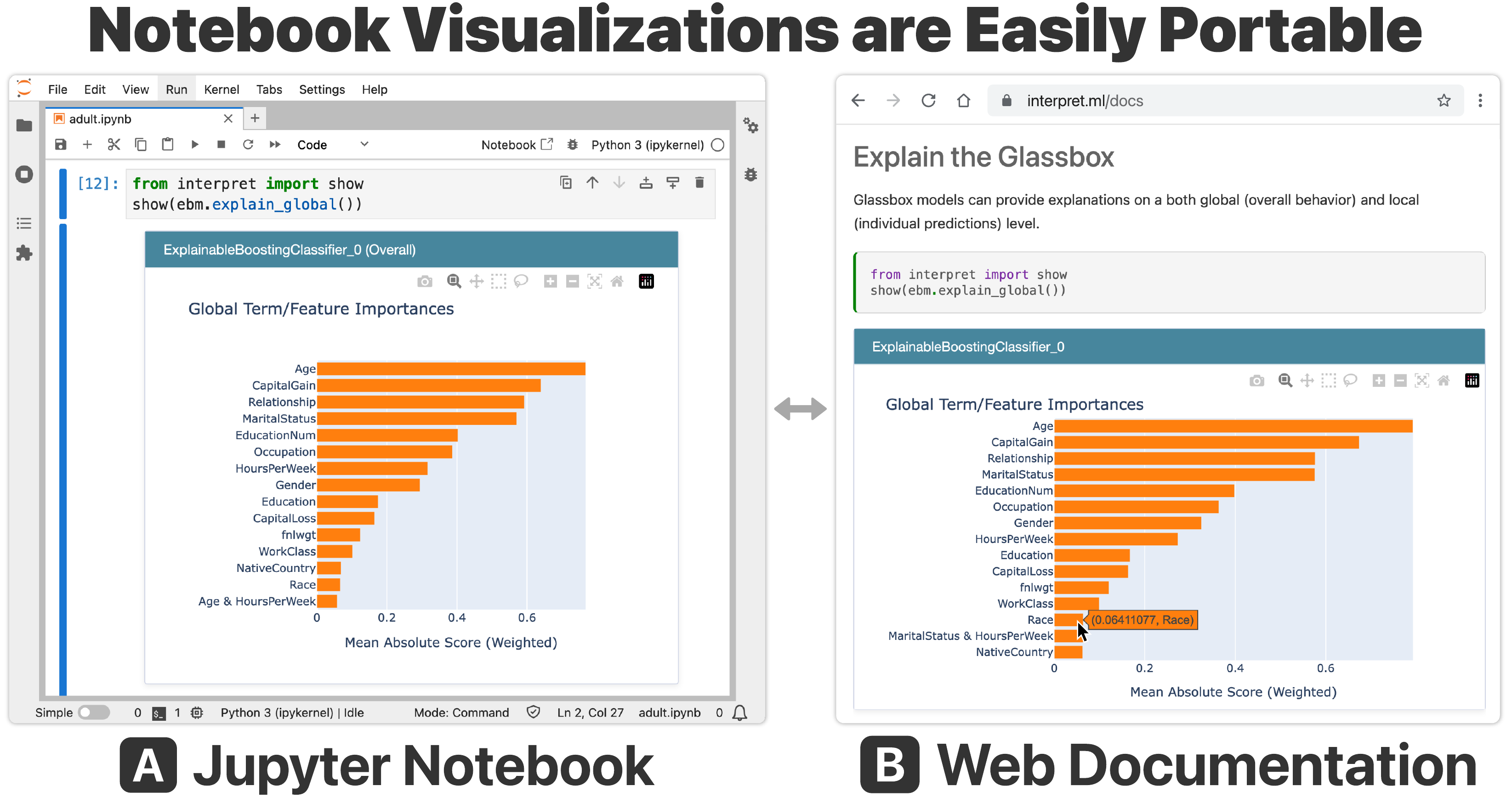}
  \caption{
    The vibrant notebook ecosystem enables developers to easily transfer their visualizations across various platforms.
    For example, (A) the Python library \package{InterpretML}~\cite{wangInterpretabilityThenWhat2022}'s notebook explainable ML visualizations are also used on (B) its documentation website via Jupyter Book~\cite{executablebookscommunityJupyterBook2020}.
  }
  \label{fig:portability}
  \Description{Screenshots of the package InterpretML and its documentation website.}
\end{figure}

The notebook community has developed a vibrant ecosystem to convert notebooks into a wide range of mediums.
This includes the ability for users to publish notebooks containing interactive visualizations as slides~\cite{wangSlide4NCreatingPresentation2023}, interactive books~\cite{executablebookscommunityJupyterBook2020}, and dashboards~\cite{bauerleSymphonyComposingInteractive2022}.
Therefore, given the portability of notebooks, notebook visualization tools have the potential to reach a more diverse audience.
For instance, \package{InterpretML}~\cite{noriInterpretMLUnifiedFramework2019} leverages Jupyter Book~\cite{executablebookscommunityJupyterBook2020} to incorporate in-notebook visualizations into its documentation, providing readers with an engaging way to learn about ML model explanations~(\autoref{fig:portability}).
However, different visualization modalities may present unique design challenges, such as potential accessibility concerns for interactive visualizations in presentation slides~\cite{yipVisionaryCaptionImproving2021} and the need to consider social contexts for dashboard design~\cite{sarikayaWhatWeTalk2019}.
Thus, it is crucial for researchers to carefully consider specific design constraints associated with different modalities if they decide to use notebooks as a bridge to other visualization mediums.

\aptLtoX{\begin{framed}\noindent\textbf{Implication on cross-modality design}:
  The notebook ecosystem offers various options for distributing and sharing notebook visualization tools with diverse stakeholders through various modalities~(e.g., interactive books, slides, dashboards).
  However, researchers need to consider unique design challenges associated with the targeted modalities.\end{framed}}{\columnbox{
{\xlabel[Implecation: domain-specific design]{opp:user-group}}
  \textbf{Implication on cross-modality design}:
  The notebook ecosystem offers various options for distributing and sharing notebook visualization tools with diverse stakeholders through various modalities~(e.g., interactive books, slides, dashboards).
  However, researchers need to consider unique design challenges associated with the targeted modalities.}}

\setlength{\belowcaptionskip}{-5pt}
\setlength{\abovecaptionskip}{5pt}
\setlength{\belowcaptionskip}{0pt}
\setlength{\abovecaptionskip}{10pt}

\subsection{Ease of Implementation}
\label{sec:why:ease}
\headerbottomspace{}

There exist multiple methods, varying in difficulty, for implementing notebook visualization tools.
Some methods are simple and attract researchers to add notebook support for existing visualizations.
For example, the ML library \package{CatBoost}~\cite{prokhorenkovaCatBoostUnbiasedBoosting2019} uses Jupyter Notebook's native \texttt{ipywidgets} to add checkboxes and sliders to help users customize simple loss function plots.
Recent researchers have introduced NOVA workflow~\cite{wangNOVAPracticalMethod2022a}, which enables easy conversion of web-based visualization apps into notebook widgets~\cite[e.g.,][]{wangTimberTrekExploringCurating2022a,munechikaVisualAuditorInteractive2022,wangInterpretabilityThenWhat2022a}.
Moreover, we observe that some developers use notebooks as a platform for rapidly prototyping and deploying GUI applications.
For instance, \package{Pigeon}~\cite{germanidisPigeonQuicklyAnnotate2017} leverages \texttt{ipywidgets} to implement a simple visualization tool that allows annotators to label text and image data.
Computational notebooks are web-based systems, and the low barrier to authoring notebook visualization tools reflects and contributes to the trend of web-based interactive visualizations~\cite{battleBeagleAutomatedExtraction2018,battleExploringD3Implementation2022}.
With the increasing ease of developing notebook visualization tools, we anticipate a growing number of such tools catering to various notebook user groups~(\autoref{sec:why:workflow}).

\aptLtoX{\begin{framed}\noindent\textbf{Implication on growing trend of notebook visualization tools}:
  As the implementation is becoming increasingly accessible, the trend of using computational notebooks as a flexible platform for deploying and developing web-based interactive visualization tools will continue.\end{framed}}{\columnbox{
{\xlabel[Implecation: growing trend]{opp:ease}}
  \textbf{Implication on growing trend of notebook visualization tools}:
  As the implementation is becoming increasingly accessible, the trend of using computational notebooks as a flexible platform for deploying and developing web-based interactive visualization tools will continue.}}
\setlength{\belowcaptionskip}{-5pt}
\setlength{\abovecaptionskip}{2pt}
\begin{figure*}[tb]
  \includegraphics[width=\linewidth]{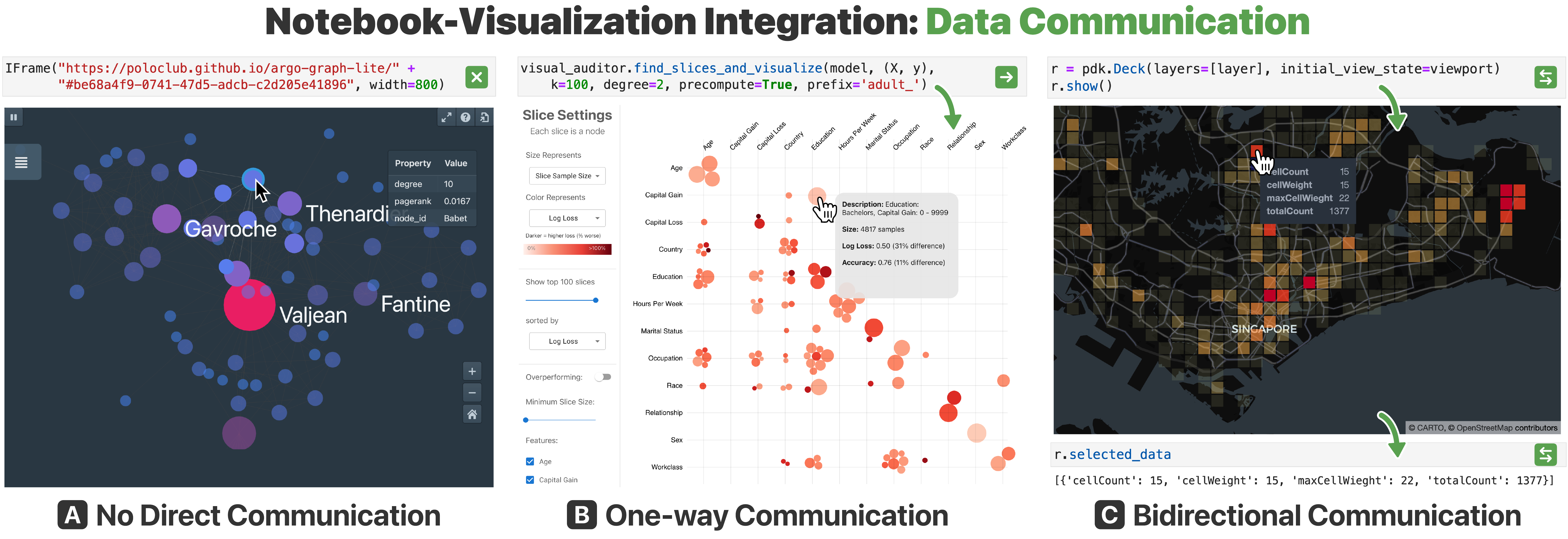}
  \caption{
    The integration level between notebook and visualization tools varies based on data communication channels.
    (A) Tools such as \package{Argo Lite}~\cite{liArgoLiteOpenSource2020} retrieve data from external servers instead of the notebook.
    (B) \package{Visual Auditor}~\cite{munechikaVisualAuditorInteractive2022} visualizes different slices of the dataset that are sent from the notebook.
    (C) More integrated tools like \package{pydeck}~\cite{uberDeckGlWebGL22016} not only visualize data from the notebook but also send data back to the notebook, for example, information on a user's selected map cells.
  }
  \label{fig:communication}
  \Description{Screenshots of Argo Lite, Visual Auditor, and Pydeck.}
\end{figure*}
\setlength{\belowcaptionskip}{0pt}
\setlength{\abovecaptionskip}{10pt}

\headertopspace{}
\section{How to Design Notebook Vis Tools}
\label{sec:design}
\headerbottomspace{}

This section discusses the design patterns of existing notebook visualization tools.
To organize these patterns, we construct a four-dimensional design space based on the tool users' needs.

\headertopspace{}
\subsection{\communicationc{Notebook-Visulization Integration}}
\label{sec:design:communication}
\headerbottomspace{}

The level of integration between notebook environments and visualization tools can vary widely.
We characterize this integration continuum by the \textit{data communication channels} between these two parties, where loosely integrated visualization tools have fewer communication channels than more tightly integrated tools.

\mypar{\vcenteredhbox{\includegraphics[height=8pt]{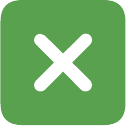}}~\communicationc{No Direct Communication.}}
A few notebook visualization tools do not directly receive data from the notebook environment, as their data source is not available within users' notebooks.
Nevertheless, notebooks allow these tools to retrieve data from external sources~(\autoref{sec:appendix:implementation:comm-no}), thereby allowing users to enjoy these tools in their workflows.
For example, \package{TensorBoard} reads log files from the file system, and \package{StatCast}~\cite{lageStatCastDashboardExploration2016} reads data from a separate database server.
\package{Argo Lite}~\cite{liArgoLiteOpenSource2020} allows notebook users to view graph visualizations that are created from a separate website~(\autoref{fig:communication}\figpart{A}).

\mypar{\vcenteredhbox{\includegraphics[height=8pt]{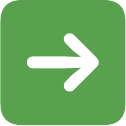}}~\communicationc{One-way Communication.}}
Most notebook visualization tools have a one-way communication with the notebook environment: they receive input from the notebook but do not send data back to the notebook~(\autoref{sec:appendix:implementation:comm-one}).
(1) Users can \textit{explicitly} specify the input.
For example, users can write code to feed an ML model and data into \package{Visual Auditor}~\cite{munechikaVisualAuditorInteractive2022}, which generates interactive visualizations for auditing model biases~(\autoref{fig:communication}\figpart{B}).
(2) Some tools also leverage \textit{implicit} input.
For instance, \package{Solas} provides situated visualization recommendations by analyzing a user's historical analysis code.
With a one-way communication, users can follow the familiar input-output notebook pattern~\cite{kluyverJupyterNotebooksPublishing2016} to customize visualization tools.

\mypar{\vcenteredhbox{\includegraphics[height=8pt]{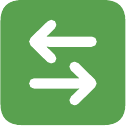}}~\communicationc{Bidirectional Communication.}}
Tools with high notebook integration not only receive input from the notebook but also update its content~(\autoref{sec:appendix:implementation:comm-two}).
(1) These tools can \textit{add new code or text} to the notebook.
For example, \package{B2}~\cite{wuB2BridgingCode2020} adds a user's interaction history to the notebook cells, and \package{Mage}~\cite{keryMageFluidMoves2020} generates code that can lead to the same consequence as user interactions.
(2) Some tools directly \textit{modify the runtime states} in a notebook.
For instance, the spatial visualization tool \package{pydeck}~\cite{uberDeckGlWebGL22016} stores the user's selected data from the visualization in a runtime variable, which users can access in other code cells~(\autoref{fig:communication}\figpart{C}).
Bidirectional communication in notebooks can be a powerful and unique feature that help interactive visualization users operationalize visualization insights~(\autoref{sec:why:artifact}).

Notebooks enable researchers to integrate both \textit{input}~(\vcenteredhbox{\includegraphics[height=8pt]{figures/icon-comm-one-c}}~\communicationc{\textbf{one-way communication}}) and \textit{output}~(\vcenteredhbox{\includegraphics[height=8pt]{figures/icon-comm-two-c}}~\communicationc{\textbf{bidirectional communication}}) of a visualization pipeline into the users' existing workflow~(\aptLtoX[graphic=no,type=html]{Implecation: new opportunities}{\nameref{opp:artifact}}).
However, designing \vcenteredhbox{\includegraphics[height=8pt]{figures/icon-comm-two-c}}~\communicationc{\textbf{bidirectional communication}} requires caution.
\citet{chattopadhyayWhatWrongComputational2020} find that notebook users often struggle to keep track of the states in different cells.
Therefore, automatically modifying notebook states through a visualization tool could cause further confusion.
Similarly, in \citet{wuB2BridgingCode2020}'s study, some participants found it "annoying" when notebook content was populated from a visualization tool.
Thus, it is crucial to offer users clear feedback and allow users to configure state-updating behaviors.

\aptLtoX{\begin{framed}\noindent\textbf{Trade-off on data communication}:
  Designing data communication channels (e.g., one-way vs. bidirectional communication) in notebook visualization tools requires careful balance: while bidirectional communication enriches user workflow, it also risks confusion, highlighting the need for clear user feedback and configurable 
content update policies.\end{framed}}{\columnbox{
{\xlabel[Trade-off: data communicaiton]{opp:communication}}
  \textbf{Trade-off on data communication}:
  Designing data communication channels (e.g., one-way vs. bidirectional communication) in notebook visualization tools requires careful balance: while bidirectional communication enriches user workflow, it also risks confusion, highlighting the need for clear user feedback and configurable 
content update policies.}}

\subsection{\datac{Data Source and Type}}
\label{sec:design:data}

Notebook environments offer rich and multimodal data sources that a visualization tool can use to meet user needs.

\mypar{\vcenteredhbox{\includegraphics[height=8pt]{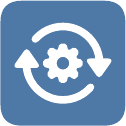}}~\datac{Runtime Artifacts.}}
The most common visualization data source is a notebook's runtime artifacts.
Visualization tools have access to any data specified by notebook users; existing notebook visualization tools support many data modalities, such as tables~\cite{brugmanPandasprofilingExploratoryData2019}, spatial data~\cite{uberDeckGlWebGL22016}, and 3D images~\cite{abrahamMachineLearningNeuroimaging2014}.
Some tools also leverage ML models in a notebook runtime, helping users interpret transformers~\cite{vigMultiscaleVisualizationAttention2019}, curate decision trees~\cite{wangTimberTrekExploringCurating2022a}, calibrate generalized additive models~\cite{xenopoulosCalibrateInteractiveAnalysis2023}, and explore counterfactual explanations~\cite{wexlerWhatIfToolInteractive2019}.

\mypar{\vcenteredhbox{\includegraphics[height=8pt]{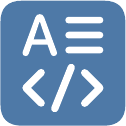}}~\datac{Code and Text.}}
Notebooks combine code and text documentation, which visualization tools can exploit to enhance visualizations~(\autoref{sec:appendix:implementation:code}).
For example, \package{Anteater}~\cite{faustInteractiveVisualizationData2022} leverages trace-based visualization to help notebook users debug their analysis code.
\package{Jigsaw}~\cite{kluyverJupyterNotebooksPublishing2016} uses variable names in a notebook to validate and correct code generated by AI models.
More recently, researchers also use code and text in notebooks to create interactive slides to communicate data insights~\cite{wangSlide4NCreatingPresentation2023,liNotableOntheflyAssistant2023}.
Moreover, to help users write high-quality ML model documentation, \package{DocML}~\cite{bhatAspirationsPracticeModel2023} links a model card visualization to both code and text cells in a notebook.

\mypar{\vcenteredhbox{\includegraphics[height=8pt]{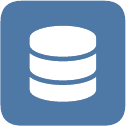}}~\datac{External Data.}}
Moreover, notebook visualization tools can access data beyond the notebook environment, such as the file system, networks, and hardware information.
For example, \package{TensorBoard} and \package{StatCast} visualize data from a local directory and a database server, respectively.
\package{NVDashboard}~\cite{nvidiaNVDashboardJupyterLabExtension2021} provides notebook users with an interactive dashboard to monitor real-time GPU usage.

Although notebooks provide unique and valuable data for designing interactive visualization tools, accessing various data types requires different implementation strategies.
While it is relatively easy to read \vcenteredhbox{\includegraphics[height=8pt]{figures/icon-data-runtime-c}}~\datac{\textbf{runtime artifacts}}~(\autoref{sec:why:ease}), it requires more engineering effort to read \vcenteredhbox{\includegraphics[height=8pt]{figures/icon-data-text-c}}~\datac{\textbf{code and text}} or implement \vcenteredhbox{\includegraphics[height=8pt]{figures/icon-comm-two-c}}~\communicationc{\textbf{bidirectional communication}}~(see \autoref{sec:appendix:implementation} for detailed discussion on implementation strategies).
Certain strategies are only compatible with specific notebook platforms; for example, tools implemented with Jupyter extensions cannot be used in Google Colab.
Thus, there is a trade-off between accessing powerful notebook features and ensuring compatibility with diverse notebook platforms.

\aptLtoX{\begin{framed}\noindent\textbf{Trade-off on compatibility}:
  Notebooks provide access to unique data types, including runtime artifacts, code, text and external data.
  However, there is a trade-off between leveraging powerful, yet platform-specific features like reading code and text and bidirectional communication, and ensuring broader compatibility across various notebook platforms.\end{framed}}{\columnbox{
{\xlabel[Trade-off: compatibility]{opp:data}}
  \textbf{Trade-off on compatibility}:
  Notebooks provide access to unique data types, including runtime artifacts, code, text and external data.
  However, there is a trade-off between leveraging powerful, yet platform-specific features like reading code and text and bidirectional communication, and ensuring broader compatibility across various notebook platforms.}}

\subsection{\displayc{Display Style \& Sensemaking Context}}
\label{sec:design:display}

\setlength{\belowcaptionskip}{-10pt}
\setlength{\abovecaptionskip}{5pt}
\begin{figure}[tb]
  \includegraphics[width=\linewidth]{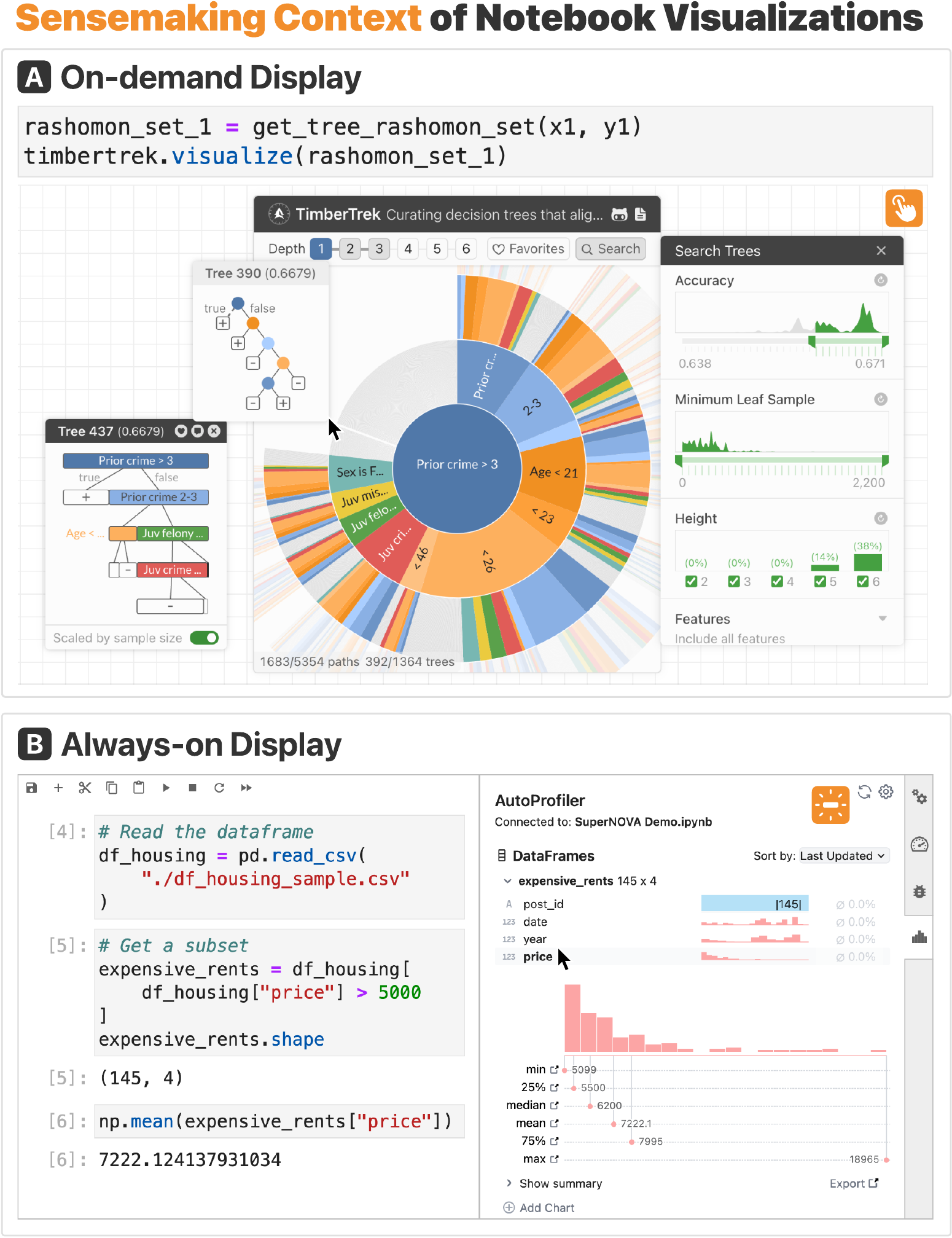}
  \caption{
    Notebook visualization tools' display styles vary based on the user's sensemaking context.
    For example, (A) \package{TimberTrek}~\cite{wangTimberTrekExploringCurating2022a} uses an on-demand display to visualize a large collection of decision trees next to the cells where the trees are created.
    (B) \package{AutoProfiler}~\cite{eppersonDeadAliveContinuous2023} leverages an always-on display to automatically and continuously highlight data distributions and summary statistics of the user's datasets.
  }
  \label{fig:display}
  \Description{Screenshots of TimberTrek and AutoProfiler.}
\end{figure}
\setlength{\belowcaptionskip}{0pt}
\setlength{\abovecaptionskip}{10pt}

Notebook visualization tools' display styles can vary based on the user's sensemaking context~\cite{liuMentalModelsVisual2010}.
On-demand displays can be used for situational contexts, while always-on displays are suitable for continuous contexts.

\mypar{\vcenteredhbox{\includegraphics[height=8pt]{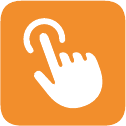}}~\displayc{On-demand display.}}
Most visualization tools show visualizations below a code cell~\cite[e.g.,][]{wexlerWhatIfToolInteractive2019,wangInterpretabilityThenWhat2022a,xenopoulosCalibrateInteractiveAnalysis2023}.
These visualizations are part of the cell flow---they move vertically with the cells when a user scrolls through the notebook.
With this layout, users can easily create multiple instances of the same visualization tool with different input data.
For example, users can create multiple instances of \package{TimberTrek}~\cite{wangTimberTrekExploringCurating2022a} in different cells with different collections of decision trees and compare across these collections~(\autoref{fig:display}\figpart{A}).

\mypar{\vcenteredhbox{\includegraphics[height=8pt]{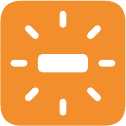}}~\displayc{Always-on display.}}
Notebook tools can also display visualizations outside of notebook cells~(\autoref{sec:appendix:implementation:always}), leading to an always-on display detached from the cell flow.
For instance, \package{AutoProfiler}~\cite{wuB2BridgingCode2020} continuously updates data distribution visualizations in a resizable dashboard pane to the right of the notebook UI, allowing users to view persistent data profiling information while exploring their datasets~(\autoref{fig:display}\figpart{B}).
Similarly, \package{NVDashboard}~\cite{nvidiaNVDashboardJupyterLabExtension2021} displays multiple charts outside of the notebook UI so that users can monitor their GPU usage in real time while interacting with the notebook.

The design choice of visualization display style in the notebook depends on the users' needs and the sensemaking context.
Based on \citet{liuMentalModelsVisual2010}'s sensemaking model, visualizations provide external anchoring, cognitive offloading and information foraging.
They suggest that visualization designers should minimize the ``semantic distance''~\cite{hutchinsDirectManipulationInterfaces1985} between the tasks users want to perform and the physical form of visualizations.
In computational notebooks, an \vcenteredhbox{\includegraphics[height=8pt]{figures/icon-display-demand-c}}~\displayc{\textbf{on-demand display}} can assist users with situational sensemaking and temporary anchoring for comparisons.
On the other hand, an \vcenteredhbox{\includegraphics[height=8pt]{figures/icon-display-always-c}}~\displayc{\textbf{always-on display}} can be beneficial for ongoing monitoring and tasks that require continuous cognitive offloading.

\aptLtoX{\begin{framed}\noindent\textbf{Trade-off on display style}:
  Researchers need to consider the trade-off between on-demand and always-on displays of interactive visualizations in notebooks based on the users' needs.
  On-demand displays aid situational sensemaking and comparisons, while always-on displays support continuous monitoring and cognitive offloading.\end{framed}}{\columnbox{
{\xlabel[Trade-off: display style]{opp:display}}
\textbf{Trade-off on display style}:
  Researchers need to consider the trade-off between on-demand and always-on displays of interactive visualizations in notebooks based on the users' needs.
  On-demand displays aid situational sensemaking and comparisons, while always-on displays support continuous monitoring and cognitive offloading.}}

\subsection{\modularityc{Modularity}}
\label{sec:design:modularity}

\noindent{}Modularity in notebook visualization tools is a critical consideration when catering to different analysis needs, such as exploratory and exploitative~\cite{batchInteractiveVisualizationGap2018}, and user's programming proficiency.
This ensures a balance between the code and the graphical user interface.

\begin{figure}[tb]
  \includegraphics[width=233pt]{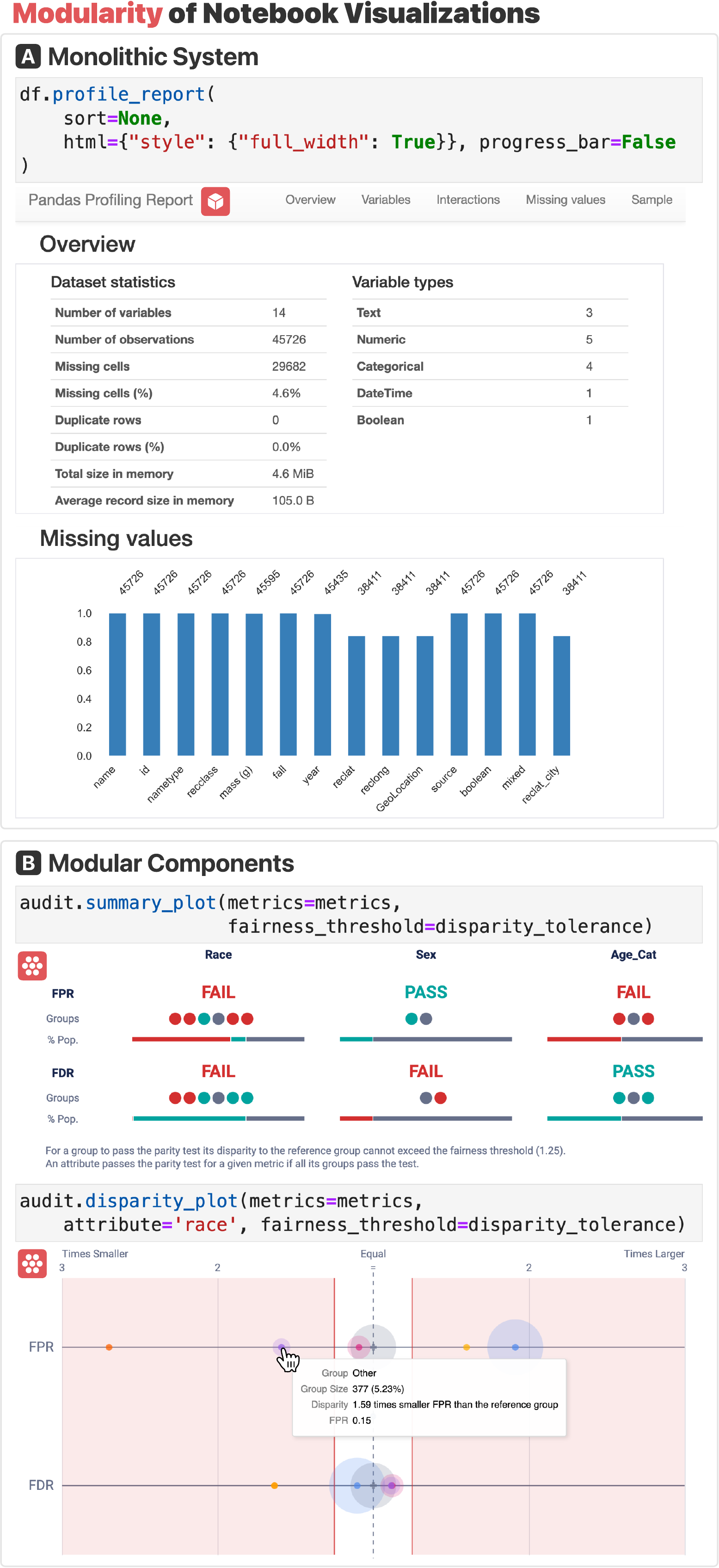}
  \caption{
    The modularity of notebook visualization tools varies.
    For example, (A) \package{ydata-profiling}~\cite{brugmanPandasprofilingExploratoryData2019} shows all components in a notebook cell, where users navigate data profiling visualizations via tabs.
    (B) In contrast, \package{Aequitas}~\cite{saleiroAequitasBiasFairness2019} modularizes different visualization components into distinct Python functions, enabling users to write code to show ML model fairness visualizations tailored to specific needs.
  }
  \label{fig:modularity}
  \Description{Screenshots of ydata-profiling and Aequitas.}
\end{figure}

\mypar{\vcenteredhbox{\includegraphics[height=8pt]{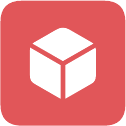}}~\modularityc{Monolithic System.}}
Most notebook visualization tools are monolithic, presenting the entire system all at once.
For example, when a user calls \package{ydata-profilling}~\cite{brugmanPandasprofilingExploratoryData2019} in a notebook cell, the tool displays a panel beneath the cell that contains all exploratory data analysis visualizations~(\autoref{fig:modularity}\figpart{A}).
These visualizations are organized into multiple tabs based on their tasks, such as variable interactions, correlations, and missing values.

\setlength{\belowcaptionskip}{-5pt}
\setlength{\abovecaptionskip}{3pt}
\setlength{\belowcaptionskip}{0pt}
\setlength{\abovecaptionskip}{10pt}

\mypar{\vcenteredhbox{\includegraphics[height=8pt]{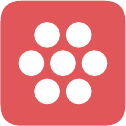}}~\modularityc{Modular Components.}}
Modular visualization tools accommodate the fragmentary nature of notebook code and allow users to easily \textit{customize} and \textit{compose} visualizations.
For example, \package{Aequitas}~\cite{saleiroAequitasBiasFairness2019}, an ML fairness auditing toolkit, provides different interactive visualizations for different fairness metrics.
These visualizations are modularized into separate functions, enabling users to write code to generate and compose visualizations that meet specific needs~(\autoref{fig:modularity}\figpart{B}).
For instance, with \package{Aequitas}, a user can create and inspect a fairness overview in a notebook cell and delve into specific fairness metrics in other separate cells.

Monolithic and modular architectures have been extensively discussed in the software engineering literature for decades~\cite{aoyamaAgileSoftwareProcess1998}.
Within the visual analytics research community, there is a recent trend towards shifting from designing ``over-complicated'' monolithic systems to simpler and reusable modular components~\cite{wuDefenceVisualAnalytics2022, bertiniBuildingEasyToAdoptSoftware2022}.
The use of \vcenteredhbox{\includegraphics[height=8pt]{figures/icon-modular-c}}~\modularityc{\textbf{modular components}} aligns well with computational notebooks, as notebook users can easily display and customize different components in separate notebook cells.
Additionally, users can take advantage of dashboard authoring tools~\cite{wangStickyLandBreakingLinear2022,bauerleSymphonyComposingInteractive2022} to compose different visualization components into a dashboard directly in their notebooks~(\autoref{sec:why:portability}).
However, \vcenteredhbox{\includegraphics[height=8pt]{figures/icon-modular-c}}~\modularityc{\textbf{modular components}} require the users to know their visualization goals~(i.e., \textit{exploitative} analysis) and know how to write code to display the appropriate components.
In contrast, a \vcenteredhbox{\includegraphics[height=8pt]{figures/icon-mono-c}}~\modularityc{\textbf{monolithic system}} is more friendly to beginner users and suitable for \textit{exploratory} analysis, where it can guide users to uncover data patterns and insights.

\aptLtoX{\begin{framed}\noindent\textbf{Trade-off on modularity}:
  Modular visualization tools are composable and reusable, particularly in notebooks where users can easily display and customize them.
  While modular components offer flexibility for users with clear analysis goals and coding skills, monolithic systems remain more beginner-friendly and ideal for exploratory analysis.\end{framed}}{\columnbox{
{\xlabel[Trade-off: modularity]{opp:modularity}}
  \textbf{Trade-off on modularity}:
  Modular visualization tools are composable and reusable, particularly in notebooks where users can easily display and customize them.
  While modular components offer flexibility for users with clear analysis goals and coding skills, monolithic systems remain more beginner-friendly and ideal for exploratory analysis.}}

\headertopspace{}
\section{Analysis}
\label{sec:evaluation}
\headerbottomspace{}

Leveraging our organizational framework as a lens, we conduct a quantitative analysis to study the relationship between the design of notebook visualization tools and the impacts of these tools (e.g., GitHub star count and publication citation count).
Our analysis offers additional insights into future design decisions.

\mypar{Data Collection.}
We characterize all \totalcount{} notebook visualization tools using our framework~(Table 1).
Then, we collect the GitHub star count, first commit date, publication year, and citation count of

\setlength{\belowcaptionskip}{-2pt}
\setlength{\abovecaptionskip}{3pt}
\begin{figure*}[tb]
  \includegraphics[width=\linewidth]{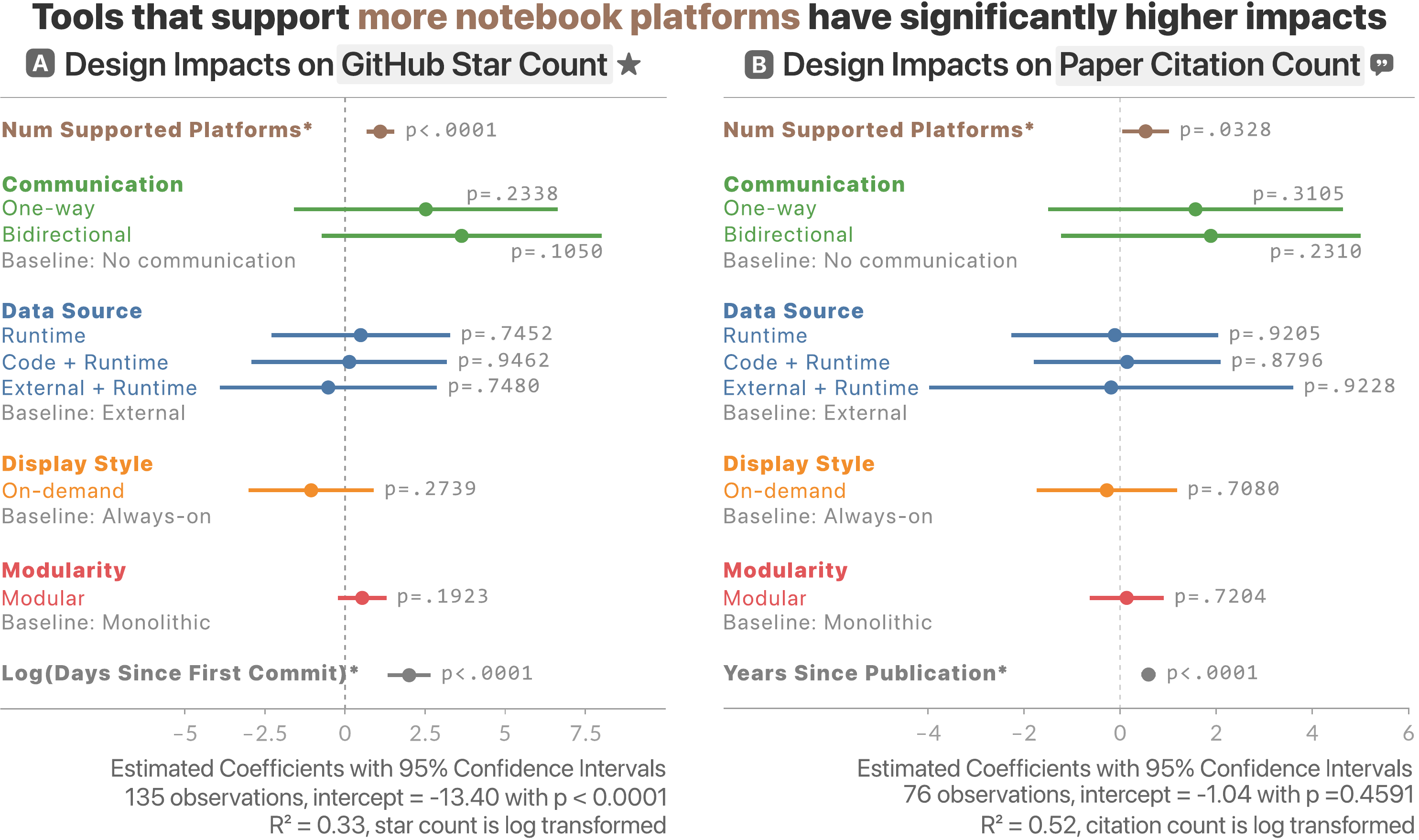}
  \caption{
    We conduct two regression analyses to investigate the effects of various design factors on the impact of notebook visualization tools, as measured by \textbf{(A)} GitHub star count and \textbf{(B)} paper citation count.
    We encode categorical variables using dummy variables, with the baseline category labeled in the figure.
    The results reveal that, in addition to time, notebook interactive visualization tools that support more notebook platforms have significantly more GitHub stars and paper citations.
  }
  \Description{This is a comparative figure divided into two panels: A and B. Both panels analyze the impacts of various design features on the success of notebook platforms, with panel A focusing on GitHub Star Count and panel B on Paper Citation Count.

    Panel A, titled "Design Impacts on GitHub Star Count," shows a horizontal axis measuring the estimated coefficients with 95\% confidence intervals. Key variables such as "Communication," "Data Source," "Display Style," and "Modularity" are compared against baselines. Notable is the "Num Supported Platforms," with a statistically significant p-value (p < 0.0001). The graph indicates that bidirectional communication, integrating code with runtime as a data source, on-demand display style, and modular design have positive impacts on GitHub star counts.

    Panel B, titled "Design Impacts on Paper Citation Count," mirrors the design of panel A but focuses on the number of citations a paper receives. Again, the "Num Supported Platforms" is statistically significant (p < 0.0001). Bidirectional communication, integration of code and runtime, on-demand display, and modularity are shown with their respective impacts on citation counts.

    Both panels include a note indicating that the log of days since the first commit or years since publication is significantly related to the outcome. The bottom of each panel reports the number of observations, the intercept, the significance of the intercept, and the R-squared value, indicating the proportion of variance explained by the models. The star and quotation mark icons next to the titles of the panels symbolize GitHub stars and citations, respectively.}
  \label{fig:coefficient}
\end{figure*}
\setlength{\belowcaptionskip}{0pt}
\setlength{\abovecaptionskip}{12pt}

\setlength{\belowcaptionskip}{-8pt}
\setlength{\abovecaptionskip}{2pt}
\begin{figure}[tb]
  \includegraphics[width=\linewidth]{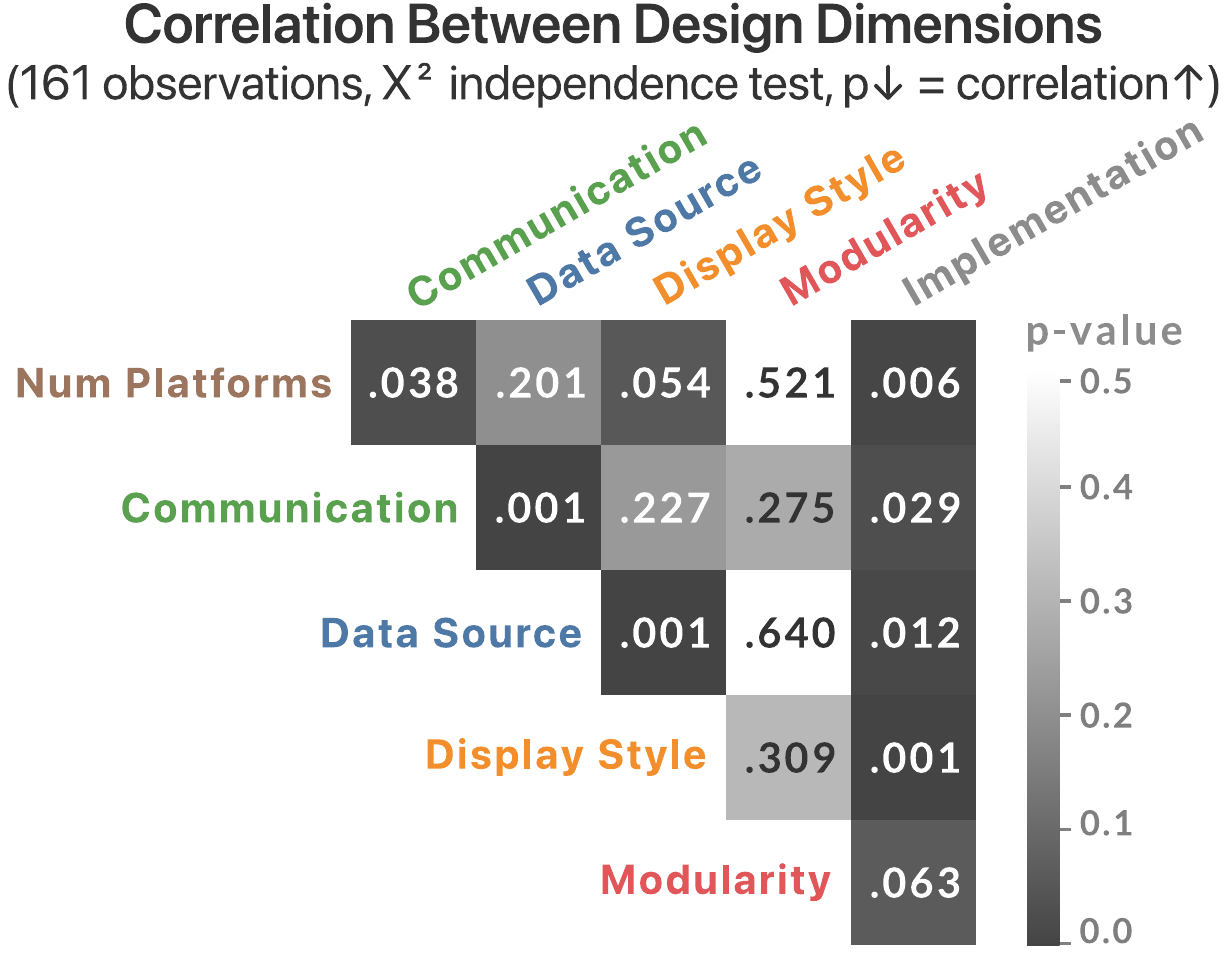}
  \caption{
    Design dimension correlations via pair-wise $X^2$ tests.
  }
  \Description{
    This figure is a matrix-style figure titled "Correlation Between Design Dimensions." It presents a table with p-values that indicate the strength and significance of correlations between various design dimensions for notebook platforms, based on 161 observations using the Chi-squared independence test.

    The design dimensions listed both horizontally and vertically include "Num Platforms," "Communication," "Data Source," "Display Style," "Modularity," and "Implementation." Each cell in the matrix provides the p-value of the correlation between the dimensions intersecting at that cell, with values ranging from 0.001 to 0.640. Lower p-values suggest stronger evidence against the null hypothesis of no association.

    The color gradient scale on the right side of the figure ranges from light to dark grey, corresponding to p-values from 0.0 to -0.5, indicating that darker cells have a lower p-value and thus a stronger correlation. For instance, the correlation between "Num Platforms" and "Modularity" has a p-value of 0.006, indicating a significant correlation, while "Data Source" and "Display Style" have a p-value of 0.640, suggesting a weaker correlation.
  }
  \label{fig:correlation}
\end{figure}
\setlength{\belowcaptionskip}{0pt}
\setlength{\abovecaptionskip}{10pt}

\noindent{}all available tools via the GitHub API~\cite{githubGitHubGraphQLAPI2023} and Semantic Scholar API~\cite{kinneySemanticScholarOpen2023}.
Among all \totalcount{} tools, 135 have GitHub repositories and 76 have Semantic Scholar entries.

\mypar{Correlation Analysis.}
We analyze the correlations across different design dimensions by conducting pair-wise $X^2$ independence tests~(\autoref{fig:correlation}).
Unsurprisingly, our results highlight that \implementationc{\textbf{implementation strategies}} correlate with many other design dimensions.
For example, tools that support \vcenteredhbox{\includegraphics[height=8pt]{figures/icon-display-always-c}}~\displayc{\textbf{always-on display}} are more likely to be \implementationc{\textbf{implemented using} \texttt{extensions}}.
Interestingly, \datac{\textbf{data source}} is correlated with both \communicationc{\textbf{communication}} and \displayc{\textbf{display}} styles.
In particular, tools that access \vcenteredhbox{\includegraphics[height=8pt]{figures/icon-data-text-c}}~\datac{\textbf{code and text}} from the notebook are more likely to support \vcenteredhbox{\includegraphics[height=8pt]{figures/icon-comm-two-c}}~\communicationc{\textbf{bidirectional communication}} and \displayc{\vcenteredhbox{\includegraphics[height=8pt]{figures/icon-display-always-c}}~\textbf{always-on display}}.
We hypothesize that this is because designers often use text and code from a notebook for generative tasks (e.g., automatic visualization generation), and they prefer always-on displays to provide notebook users with continuous feedback~(\aptLtoX[graphic=no,type=html]{Trade-off: display style}{\nameref{opp:display}}).
For example, visualization recommendation tools B2~\cite{wuB2BridgingCode2020} and PI2~\cite{chenPI2EndtoendInteractive2022} leverage existing code and text in a notebook to generate new visualization code in the notebook and display synthesized visualizations on an always-on panel.
Finally, we observe that tools with \vcenteredhbox{\includegraphics[height=8pt]{figures/icon-comm-two-c}}~\communicationc{\textbf{bidirectional communication}} support much less \platformc{\textbf{notebook platforms}} than tools with \communicationc{\vcenteredhbox{\includegraphics[height=8pt]{figures/icon-comm-one-c}}~\textbf{one-way communication}}.
This empirical finding reflects the trade-off between notebook integration and platform dependency~(\aptLtoX[graphic=no,type=html]{Trade-off: compatibility}{\nameref{opp:data}}).

\mypar{Regression Analysis.}
We conduct two regression analyses to examine the effects of design factors on the impact of notebook visualization tools, as measured by GitHub star and paper citation counts~(\autoref{fig:coefficient}).
Since \implementationc{\textbf{implementation strategies}} correlate with many other design dimensions, we do not include it in both regression models.
We include time as an independent variable and use dummy variables to encode categorical variables.
The results highlight that tools \platformc{\textbf{supporting more notebook platforms}} have significantly more GitHub stars and paper citations.
Other design dimensions do not significantly affect the popularity and recognition of notebook visualization tools.
This result implies that future researchers and developers should prioritize notebook platform compatibility to maximize the impact of their tools.

\headertopspace{}
\section{Discussion and Future Work}
\headerbottomspace{}

By analyzing \totalcount{} interactive notebook visualization tools identified from \notebookcount{} public notebooks and \papercount{} academic papers~(\autoref{sec:method}), we present an organization framework to characterize these tools~(\autoref{sec:why}, \autoref{sec:design}).
We provide practice design implications and trade-offs as well as insights from statistical analyses~(\autoref{sec:evaluation}).
Based on our findings, we discuss future research opportunities and limitations of our study.

\mypar{Democratizing Notebook Visualization Tool Creation.}
We discover a spectrum of methods, varying in difficulty, for authoring notebook visualization tools~(\autoref{sec:why:ease}).
In particular, accessing \vcenteredhbox{\includegraphics[height=8pt]{figures/icon-data-text-c}}~\datac{\textbf{code and text}} and supporting \vcenteredhbox{\includegraphics[height=8pt]{figures/icon-comm-two-c}}~\communicationc{\textbf{bidirectional communication}} require significant engineering effort~(\autoref{sec:design:data}).
Furthermore, some implementation strategies are only compatible with specific notebook platforms~(\aptLtoX[graphic=no,type=html]{Trade-off: compatibility}{\nameref{opp:data}}).
Therefore, we see research opportunities to lower the barrier to authoring notebook interactive visualization tools that harness the full potential of notebook platforms.
First, practitioners often use libraries such as D3~\cite{bostockDataDrivenDocuments2011} and VegaLite~\cite{satyanarayanVegaliteGrammarInteractive2017} to develop web-based interactive visualizations.
It would be valuable if these libraries integrated native support for notebook platforms or new libraries specifically targeted authoring notebook visualizations.
On the other hand, researchers can also enhance notebook platforms to better support interactive visualizations.
For example, similar to browser vendors sharing the same web standard, researchers can develop a universal notebook protocol that enables developers to access and communicate data using a standardized method across notebook platforms.

\setlength{\belowcaptionskip}{-5pt}
\setlength{\abovecaptionskip}{3pt}
\begin{figure}[tb]
  \includegraphics[width=\linewidth]{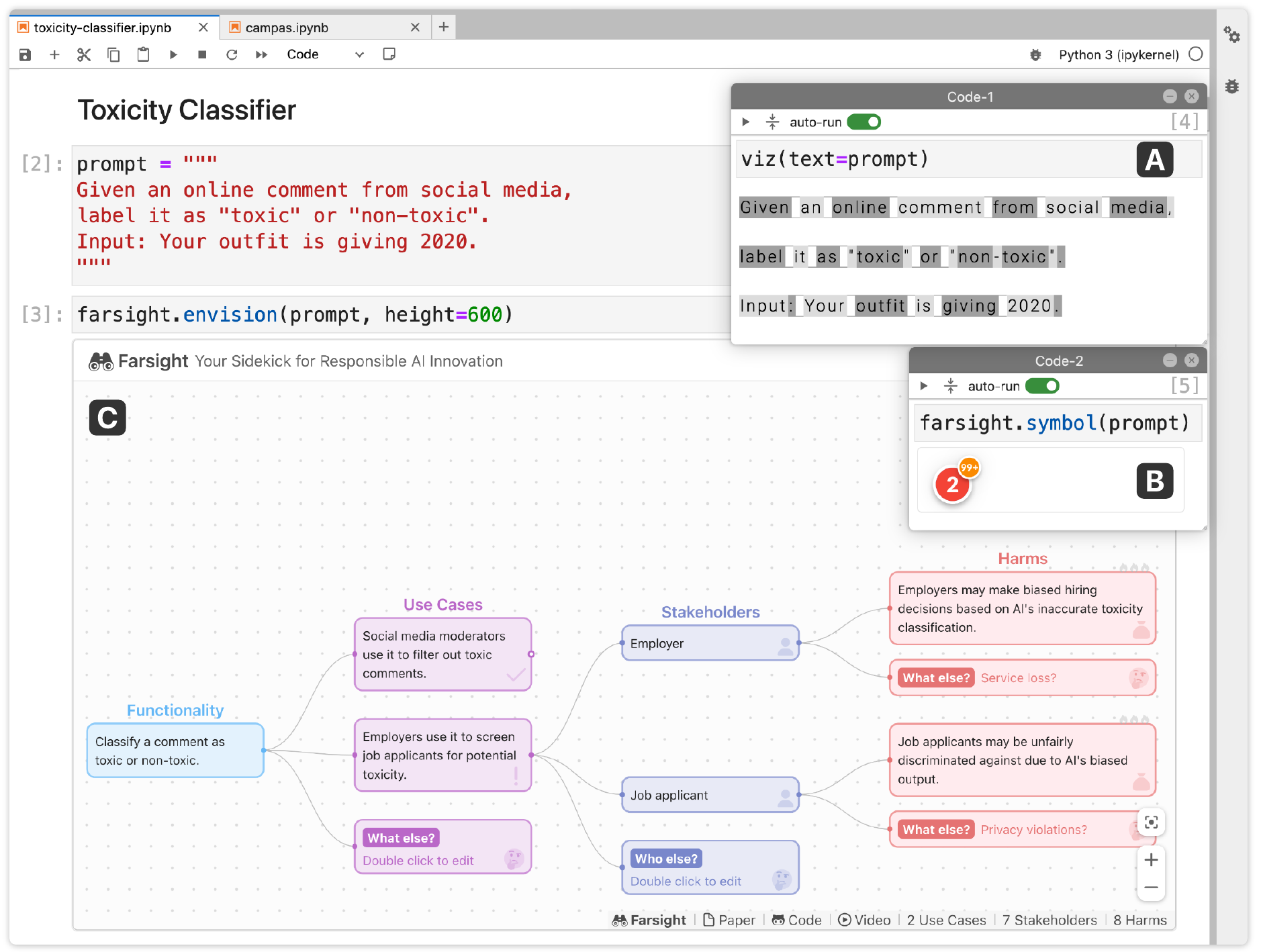}
  \caption{
    \package{StickyLand}~\cite{wangStickyLandBreakingLinear2022} enables a non-linear notebook layout, allowing notebook users to easily switch between on-demand and always-on displays of notebook visualization tools.
    For example, while prototyping large language model-powered apps by writing prompts, (A) a user can create a sticky cell with \package{HuggingFace Tokenizer}~\cite{moiHuggingFaceTokenizers2023} to continuously visualize the prompt's tokenization patterns.
    (B) The user can also use an always-on display with \package{Farsight}~\cite{wangFarsightFosteringResponsible2024} to alert to potential risks associated with the prompt.
    (C) Moreover, the user can use \package{Farsight}'s interactive tree visualization in an on-demand display for brainstorming use cases, stakeholders, and potential harms of their apps.
  }
  \label{fig:stickyland}
  \Description{Screenshots of StickyLand, Farsight, and HuggingFace Tokenizer.}
\end{figure}
\setlength{\belowcaptionskip}{0pt}
\setlength{\abovecaptionskip}{10pt}

\mypar{Enriching Fluid Notebook-Vis Integration.}
The design trade-offs regarding visualization display styles~(\aptLtoX[graphic=no,type=html]{Trade-off: display style}{\nameref{opp:display}}) and modularity~(\aptLtoX[graphic=no,type=html]{Trade-off: modularity}{\nameref{opp:modularity}}) partially arise from the rigid layout of the popular cell-based notebooks~\cite{lauDesignSpaceComputational2020}.
For example, most notebook platforms present cells in a linear manner, thereby requiring designers to decide whether to display their visualization tools within the flow~(\vcenteredhbox{\includegraphics[height=8pt]{figures/icon-display-demand-c}}~\displayc{\textbf{on-demand display}}) of the cell or detach them from the flow~(\vcenteredhbox{\includegraphics[height=8pt]{figures/icon-display-always-c}}~\displayc{\textbf{always-on display}}).
To address this trade-off, researchers can explore alternative notebook layouts.
For example, researchers have introduced sticky cells~\cite{wangStickyLandBreakingLinear2022} to break the linear presentation of notebook cells.
These sticky cells provide visualization designers with the flexibility to seamlessly switch between \vcenteredhbox{\includegraphics[height=8pt]{figures/icon-display-demand-c}}~\displayc{\textbf{on-demand}} and \vcenteredhbox{\includegraphics[height=8pt]{figures/icon-display-always-c}}~\displayc{\textbf{always-on}} displays~(\autoref{fig:stickyland}).
Similarly, regarding the modularity of visualization tools, future researchers could develop intelligent notebook interfaces that automatically adapt a visualization tool between {\vcenteredhbox{\includegraphics[height=8pt]{figures/icon-modular-c}}~\modularityc{\textbf{modular}}} and {\vcenteredhbox{\includegraphics[height=8pt]{figures/icon-mono-c}}~\modularityc{\textbf{monolithic}}} modes based on the users' current tasks and requirements.

\mypar{Promoting Responsible AI through Notebook Workflows.}
We observe an interesting trend that researchers exploit notebooks as a means to promote responsible AI practices~(e.g., \package{Aequitas}~\cite{saleiroAequitasBiasFairness2019}, \package{Fairlearn}~\cite{dudikFairlearnToolkitAssessing2020}, \package{Farsight}~\cite{wangFarsightFosteringResponsible2024}, and \package{MLDoc}~\cite{bhatAspirationsPracticeModel2023}).
We identify two motivations for this emerging trend.
First, AI practitioners often lack incentives to adopt responsible AI practices~\cite{rakovaWhereResponsibleAI2021, schiffPrinciplesPracticesResponsible2020}, such as fairness assessment and model documentation.
By integrating responsible AI practices directly into practitioners' existing notebook workflows~(\autoref{sec:why:workflow}), researchers aim to minimize adoption friction and ``nudge''~\cite{bhatAspirationsPracticeModel2023} practitioners to follow these practices.
For example, \package{Farsight} alerts users to potential harms of their large language model-powered apps while they are developing prompts in a notebook~(\autoref{fig:stickyland}).
Similarly, \package{MLDoc} automatically creates and shows an AI ``model card''~\cite{mitchellModelCardsModel2019} using content from a notebook.

Secondly, responsible AI requires collaboration across disciplines and teams within an organization~\cite{rakovaWhereResponsibleAI2021, wangDesigningResponsibleAI2023}.
Because AI practitioners have already been using notebooks to collaborate with diverse stakeholders (e.g., designers and managers)~\cite{zhangHowDataScience2020}, researchers leverage notebooks as a boundary object to facilitate responsible AI practices across teams.
For example, in \citet{dengExploringHowMachine2022}'s study on ML fairness toolkits, a participant highlighted ``\textit{a simple notebook format and compelling visualizations are needed for [organizational] leadership to adopt the toolkits.}''
Thus, as the mitigation of AI harms has become increasingly crucial, we see exciting research opportunities for researchers to design, develop, and evaluate notebook visualization tools to promote responsible AI.

\mypar{Limitations.}
In this study, to keep our review manageable and focused, we focus on computational notebooks designed for Python, the most commonly used programming language among data scientists~\cite{kaggleStateMachineLearning2022}.
Future work can explore notebooks designed for other languages, such as R Markdown~\cite{rstudioMarkdown2016} for R and Observable~\cite{observableObservableDataVisualization2021} for JavaScript.
As notebook visualization tools are still nascent, there are limited user studies evaluating the effectiveness of these tools.
In addition, although there are many different notebook user groups~(\autoref{sec:why:workflow}), the existing HCI notebook research focuses on data scientists~\cite{lauDesignSpaceComputational2020}.
To broaden the understanding of notebook visualization tools, future research endeavors can involve engaging with diverse user groups, including scientists, educators, students, and users with accessibility needs. %
\headertopspace{}
\section{Conclusion}
\headerbottomspace{}

We collect a total of \totalcount{} notebook visualization tools, including 64 from academic papers and 103 sourced from a pool of 55k notebooks containing interactive visualizations that we obtain by scraping 8.6 million notebooks on GitHub.
Based on our review, we introduce a framework for characterizing these tools in terms of their motivation for supporting notebooks, targeted users, and design patterns.
We further discuss key design implications and trade-offs as well as research opportunities for notebook visualization.
Finally, we present \tool{} to help researchers and developers easily explore existing notebook visualization tools.
We hope that our work contributes to a more comprehensive understanding of notebook visualization tools and helps researchers design and develop visualization tools that are easy to use and adopt.

\begin{acks}
  This work was supported by a J.P. Morgan PhD Fellowship, Apple Scholars in AI/ML PhD fellowship, gifts from Bosch and Cisco.
  We thank anonymous reviewers for their valuable feedback.
\end{acks}

\balance
\bibliographystyle{ACM-Reference-Format}
\bibliography{notebook-va}

\appendix

\clearpage{}
\onecolumn

\setlength{\tabcolsep}{0pt}

\aptLtoX{\normalfont
\begin{tabular}{p{41pc}}
    \includegraphics[width=\linewidth]{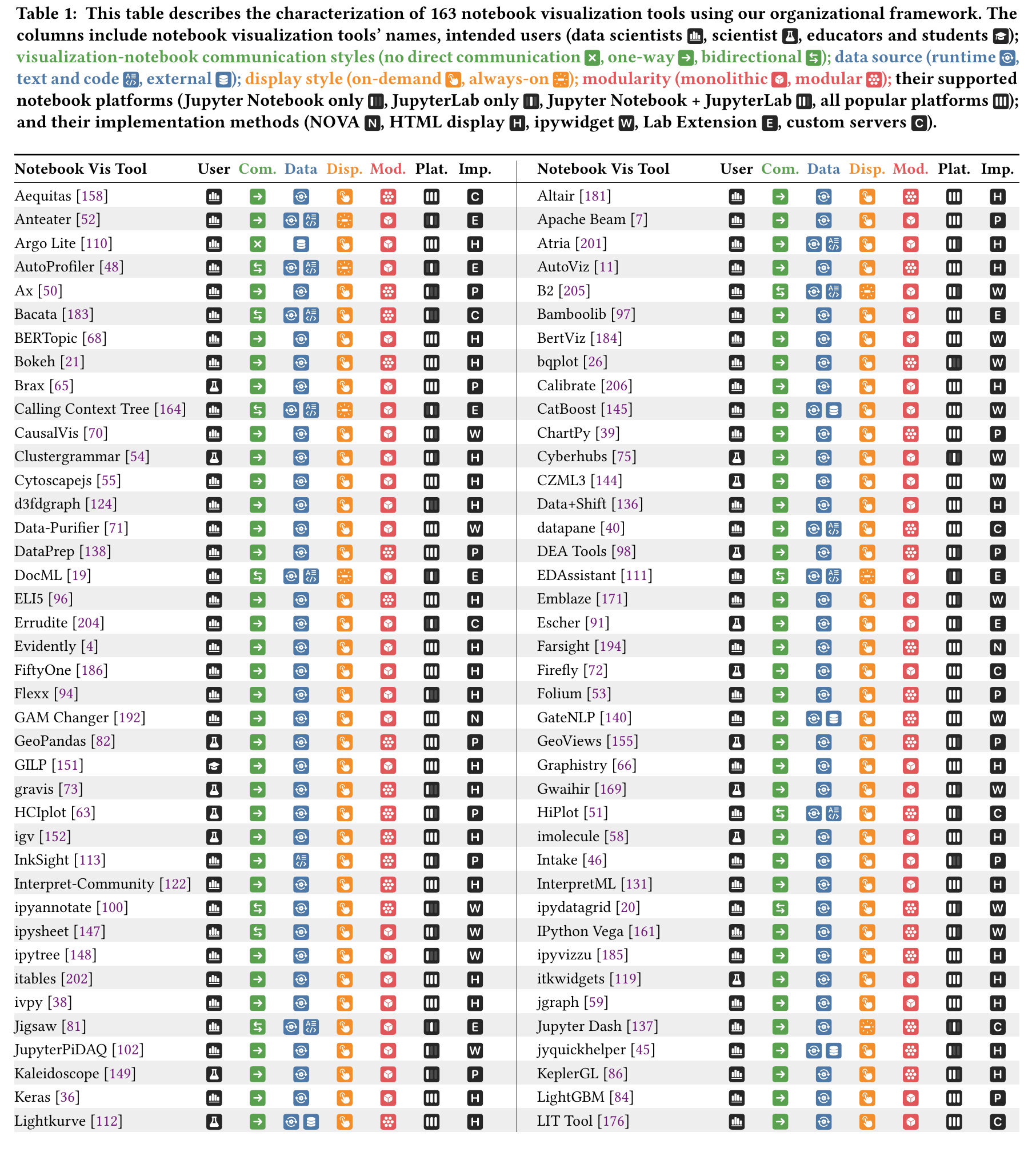} \\[-24pt]
    \includegraphics[width=\linewidth]{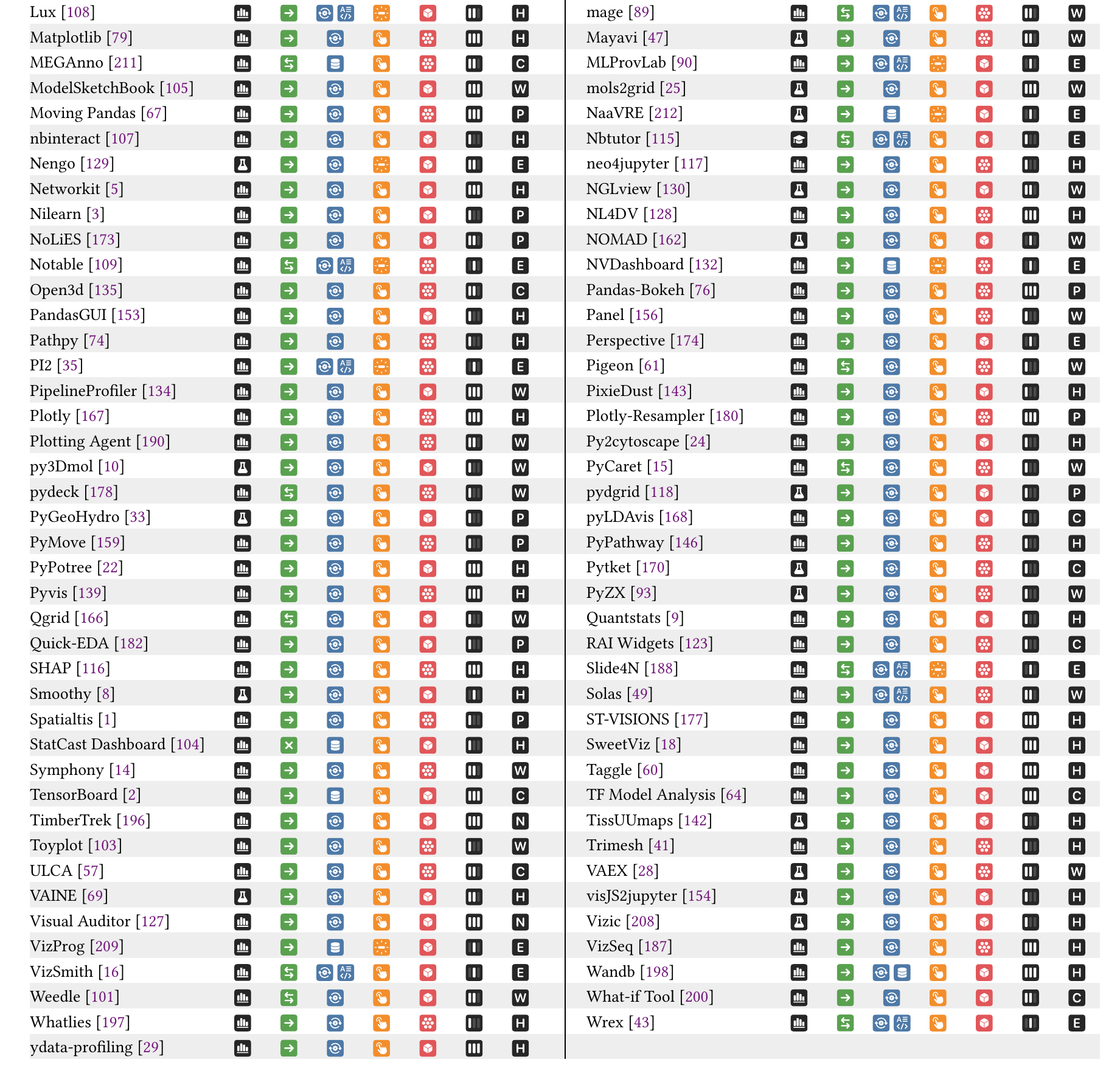}
  \end{tabular}}
{%
\section{Characterizing Notebook Visualization Tools}
\label{sec:appendixtable}
Below we characterize \totalcount{} collected notebook visualization tools using our organizational framework (targeted users and a four-dimensional design space described in~\autoref{sec:method}), as well as their supported notebook platforms and implementation methods.
See {\href{https://poloclub.github.io/supernova}{\color{blueVI}\textbf{\tool{}}}} for an interactive version with more details about each entry.

\setlength{\tabcolsep}{0pt}

\small

\begin{longtable}{
  p{90pt}
  >{\centering\arraybackslash}p{22pt}
  >{\centering\arraybackslash}p{22pt}
  >{\centering\arraybackslash}p{22pt}
  >{\centering\arraybackslash}p{22pt}
  >{\centering\arraybackslash}p{22pt}
  >{\centering\arraybackslash}p{22pt}
  >{\centering\arraybackslash}p{22pt}
  p{10pt}|
  p{10pt}
  p{90pt}
  >{\centering\arraybackslash}p{22pt}
  >{\centering\arraybackslash}p{22pt}
  >{\centering\arraybackslash}p{22pt}
  >{\centering\arraybackslash}p{22pt}
  >{\centering\arraybackslash}p{22pt}
  >{\centering\arraybackslash}p{22pt}
  >{\centering\arraybackslash}p{22pt}
  }

  \caption{\label{table:tools}
    This table describes the characterization of \totalcount{} notebook visualization tools using our organizational framework.
    The columns include notebook visualization tools' names, intended users (data scientists~\vcenteredhbox{\includegraphics[height=8pt]{figures/icon-user-data-alt}}, scientist~\vcenteredhbox{\includegraphics[height=8pt]{figures/icon-user-scientist-alt}}, educators and students~\vcenteredhbox{\includegraphics[height=8pt]{figures/icon-user-student-alt}});
    \communicationc{visualization-notebook communication styles (no direct communication~\vcenteredhbox{\includegraphics[height=8pt]{figures/icon-comm-no-c}}, one-way~\vcenteredhbox{\includegraphics[height=8pt]{figures/icon-comm-one-c}}, bidirectional~\vcenteredhbox{\includegraphics[height=8pt]{figures/icon-comm-two-c}});}
    \datac{data source (runtime~\vcenteredhbox{\includegraphics[height=8pt]{figures/icon-data-runtime-c}}, text and code~\vcenteredhbox{\includegraphics[height=8pt]{figures/icon-data-text-c}}, external~\vcenteredhbox{\includegraphics[height=8pt]{figures/icon-data-external-c}});}
    \displayc{display style (on-demand~\vcenteredhbox{\includegraphics[height=8pt]{figures/icon-display-demand-c}}, always-on~\vcenteredhbox{\includegraphics[height=8pt]{figures/icon-display-always-c}});}
    \modularityc{modularity (monolithic~\vcenteredhbox{\includegraphics[height=8pt]{figures/icon-mono-c}}, modular~\vcenteredhbox{\includegraphics[height=8pt]{figures/icon-modular-c}});}
    their supported notebook platforms (Jupyter Notebook only~\vcenteredhbox{\includegraphics[height=8pt]{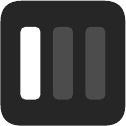}}, JupyterLab only~\vcenteredhbox{\includegraphics[height=8pt]{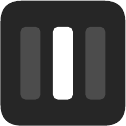}}, Jupyter Notebook + JupyterLab~\vcenteredhbox{\includegraphics[height=8pt]{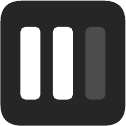}}, all popular platforms~\vcenteredhbox{\includegraphics[height=8pt]{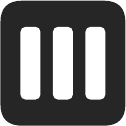}}); and their implementation methods (NOVA~\vcenteredhbox{\includegraphics[height=8pt]{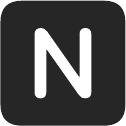}}, HTML display~\vcenteredhbox{\includegraphics[height=8pt]{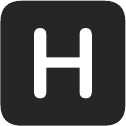}}, ipywidget~\vcenteredhbox{\includegraphics[height=8pt]{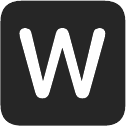}}, Lab Extension~\vcenteredhbox{\includegraphics[height=8pt]{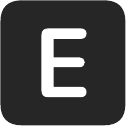}}, custom servers~\vcenteredhbox{\includegraphics[height=8pt]{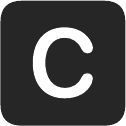}}).
  }                                                                                                                                                                                                                                                                                                                                                                                                                                                                                                                                                                                                                                                                                                                                                                                                                                  \\
\end{longtable}

\normalsize %
}

\normalsize
\twocolumn
\clearpage{}

\setcounter{figure}{0}
\renewcommand{\thetable}{S\arabic{table}}
\renewcommand{\thefigure}{S\arabic{figure}}

\section{Data Collection Details}
\label{sec:appendix:collection}

To study how researchers and practitioners design interactive visualization tools for computational notebooks, we collected and analyzed \papercount{} academic papers and \packagecount{} systems in the wild.
We define notebook visualization tools as systems that can display interactive visualizations in Python computational notebooks.

\mypar{Literature Collection.}
We searched Google Scholar for notebook visualization tools and performed forward and backward reference searches to snowball the results.
The venues of collected papers range from scientific journals (e.g., \textit{Bioinformatics} and\textit{Frontiers in Neuroinformatics}) to human-computer interaction and machine learning conferences (e.g., \textit{VIS}, \textit{CHI}, and \textit{NeurIPS}).

\mypar{Visualization Package Collection.}
We first scraped \notebookcount{} notebooks with \texttt{.ipynb} extension from GitHub.
Each notebook file is a JSON file containing metadata about the notebook and the notebook cells.
The notebook cells contain information about the cell type, the source code or text of the cell, and any output generated by the cell.
We pruned the scraped notebooks to only those containing interactive components by searching for script tags in cell outputs of type \texttt{text/html}.
If a cell was deemed a potential candidate, we extracted the associated source code for that cell.
Next, for each candidate interactive notebook, we identified all modules in the notebook by parsing it as an abstract syntax tree and looking for import statements.
Finally, we spliced the last line of the source code for the candidate cell into its individual variable components and checked if these matched any of the imported modules or their aliases.

Automating this procedure across \notebookcount{} notebooks, we built a comprehensive list of \totalpackagecount{} Python packages that were potential visualization tools.
Since this list of packages contained false positives (not all identified packages were interactive visualization tools), we manually examined each package to verify if it was an interactive visualization tool by looking at the source code and documentation for the package and its usage in notebooks.
In total, we identified \packagecount{} packages that were also interactive visualization tools.

\section{Implementation Details}
\label{sec:appendix:implementation}

Depending on the need for a backend server, visualization-notebook communication, needed data types, and display styles, there are multiple methods with varying difficulties to implement notebook visualization tools.
Note that some methods are only compatible with specific notebook platforms (e.g., JupyterLab, Colab, VSCode, and Kaggle Notebook).

\subsection{With Backend Servers}
To implement notebook visualization tools that require a backend server, the developer needs to configure the server to support notebooks and establish callback functions to share states with the notebook.
The server can either be run directly from the notebook environment or externally.
The front-end of the tool can then be displayed in the notebook using the notebook's native HTML display.
It is important to separate the server from the main thread if it is run directly from the notebook to avoid blocking the Python kernel.
For example, \package{Jupyter-Dash}~\cite{parmerDashDataApps2020} and \package{LIT}~\cite{tenneyLanguageInterpretabilityTool2020} use this method with a Flask backend server and a direct WSGI server, respectively.

\subsection{Without Backend Servers}

If the tool does not require a server, several implementation methods depend on the visualization-notebook communications.

\subsubsection{No Direct Communication.}
\label{sec:appendix:implementation:comm-no}
If a web-based visualization tool does not communicate with the notebook environment, the developer can simply use the notebook's native HTML display to show the tool a notebook cell.
The HTML display internally uses \texttt{iframe} to embed any web documents.

\subsubsection{One-way Communication.}
\label{sec:appendix:implementation:comm-one}
To pass data from the notebook Python kernel to the visualization tool, one can use the Web standard's \texttt{postMessage} method to send serialized Python objects as JSON text to the visualization tool's \texttt{iframe}.
See \package{NOVA}~\cite{wangNOVAPracticalMethod2022a} for more details and examples about this approach.
Example tools include \package{GAM Changer}~\cite{wangInterpretabilityThenWhat2022a} and \package{TimberTrek}~\cite{wangTimberTrekExploringCurating2022a}.

Alternatively, developers can use existing interactive visualization packages such as \package{Plotly}~\cite{sievertPlotlyCreateInteractive2017}, \package{Bokeh}~\cite{bokehdevelopmentteamBokehPythonLibrary2014}, \package{Altair}~\cite{vanderplasAltairInteractiveStatistical2018}, and \package{Panel}~\cite{rudigerPanelHighlevelApp2021} as building blocks to implement their visualization tools.
Then, the developer can use these packages' APIs to pass data from notebooks to the visualization tools.
However, this approach is less customizable, and it is best suited for simpler tools.
Example tools include \package{InterpretML}~\cite{noriInterpretMLUnifiedFramework2019} and \package{Nilearn}~\cite{abrahamMachineLearningNeuroimaging2014}.

\subsubsection{Bidirectional Communication.}
\label{sec:appendix:implementation:comm-two}
To send data back from the visualization tool to the Python kernel, the developer needs to use platform-specific solutions, which vary across platforms because notebook platforms have different security protocols.
For Jupyter Notebook and JupyterLab, one can use \texttt{ipywidget} with the \texttt{comm} protocol to synchronize states between the visualization tool and the notebook.
Example tools include \package{Mage}~\cite{keryMageFluidMoves2020} and \package{pydec}~\cite{uberDeckGlWebGL22016}.

\subsection{Access and Modify Code and Text}
\label{sec:appendix:implementation:code}
To access and modify notebook content outside of the Python kernel, such as raw code and text~(\autoref{sec:design:data}), visualization tool developers need to use platform-specific APIs.
For Jupyter notebooks, the developer can use \texttt{Jupyter Notebook extension} and \texttt{JupyterLab extension}  APIs to read and write the notebook content.
visualization tools using this method include \package{B2}~\cite{wuB2BridgingCode2020} and \package{Wrex}~\cite{drososWrexUnifiedProgrammingbyExample2020}.

\subsection{Always-on Display}
\label{sec:appendix:implementation:always}
If a developer intends to implement an always-on display~(\autoref{sec:design:display}) for their notebook visualization tool, they can use platform-specific APIs.
For JupyterLab, the developer can implement the tool as a \texttt{JupyterLab extension}, which enables the display on persistent panels outside of the notebook's main UI.
Examples of such implementations include \package{NVDashboard}~\cite{nvidiaNVDashboardJupyterLabExtension2021} and \package{AutoProfiler}~\cite{eppersonDeadAliveContinuous2023}.
If the visualization tool does not require extensive visualization customization, the developer can also use existing visualization packages that support persistent display (e.g., \package{Jupyter-Dash}~\cite{parmerDashDataApps2020}) to implement the tool.
Alternatively, the developer can develop their visualization tool using a traditional on-demand display and instruct users to use \package{StickyLand}~\cite{wangStickyLandBreakingLinear2022} to enable persistent display.
\package{StickyLand} allows users to easily create persistent ``sticky'' cells and dashboards by dragging any notebook cell to the edge of the notebook's UI~(\autoref{fig:stickyland}).

\end{document}